\documentclass[sigconf,screen]{acmart}
\usepackage{microtype}
\usepackage{caption}
\usepackage{multirow}
\usepackage{subcaption}
\usepackage{hyperref}
\usepackage{bbm}
\usepackage{color, colortbl}
\usepackage{xcolor}
\usepackage{tikz}
\usetikzlibrary{matrix,positioning}

\usepackage{colortbl}
\usepackage{pgfplots}
\usepackage{pgfplotstable}
\pgfplotsset{compat=1.16}
\usepackage{balance}
\usepackage{caption}
\usepackage{multirow}
\usepackage{subcaption}
\usepackage{hyperref}
\usepackage{bbm}
\usepackage{color, colortbl}
\usepackage{bigstrut}
\usepackage{booktabs}
\usepackage{footnote}
\usepackage[inline]{enumitem}
\usepackage{xspace}
\usepackage{cleveref}
\crefformat{section}{\S#2#1#3}
\crefmultiformat{section}{\S\S#2#1#3}{and~#2#1#3}{, #2#1#3}{, and~#2#1#3}
\Crefname{equation}{Eq.}{Eqs.}
\Crefname{figure}{Fig.}{Figs.}
\Crefname{tabular}{Tab.}{Tabs.}
\crefname{algocf}{pseudocode}{pseudocodes}
\Crefname{algocf}{Pseudocode}{Pseudocodes}

\makesavenoteenv{tabular}
\makesavenoteenv{table}
\definecolor{LightCyan}{rgb}{0.88,1,1}
\newcommand{\etal}{\textit{et al.}}
\newcommand{\ie}{\textit{i.e.,} }
\newcommand{\eg}{\textit{e.g.,} }
\newcommand{\subject}[1]{\texttt{\small #1}}

\newcommand{\tool}{Mono2Micro\xspace}
\newcommand{\toolnsp}{Mono2Micro\xspace}

\usepackage{boldline}
\newcolumntype{C}[1]{>{\centering\arraybackslash}p{#1}}
\definecolor{gray05}{gray}{0.95}
\definecolor{gray10}{gray}{0.90}
\definecolor{gray12}{gray}{0.88}
\definecolor{gray15}{gray}{0.85}
\definecolor{gray20}{gray}{0.80}
\definecolor{gray25}{gray}{0.75}
\definecolor{gray30}{gray}{0.70}
\definecolor{gray40}{gray}{0.60}
\definecolor{gray50}{gray}{0.50}
\definecolor{gray60}{gray}{0.40}
\definecolor{gray70}{gray}{0.30}
\definecolor{gray75}{gray}{0.25}
\definecolor{gray80}{gray}{0.20}
\definecolor{gray85}{gray}{0.15}
\definecolor{gray90}{gray}{0.10}
\definecolor{gray95}{gray}{0.05}

\usepackage[framemethod=tikz]{mdframed}
\usetikzlibrary{shadows}
\usepackage{graphics}
\newmdenv[
    tikzsetting= {fill=blue!10},
    skipabove=0.33em,
    skipbelow=0.33em,
    linewidth=1pt,
    innerleftmargin=4pt,
    innerrightmargin=4pt,
    innertopmargin=2pt,
    innerbottommargin=2pt,
    linecolor=gray85,
    roundcorner=2pt, 
    shadow=true,
    shadowsize=4pt,
    shadowcolor=black
]{myshadowbox}

\usepackage{tikz}

\newenvironment{result}
{\begin{myshadowbox}}
{\end{myshadowbox}}

\setcopyright{acmcopyright}
\acmPrice{15.00}
\acmDOI{10.1145/3468264.3473915}
\copyrightyear{2021}
\acmYear{2021}
\acmConference[ESEC/FSE '21]{Proceedings of the 29th ACM Joint European Software Engineering Conference and Symposium on the Foundations of Software Engineering}{August 23--28, 2021}{Athens, Greece}
\acmBooktitle{Proceedings of the 29th ACM Joint European Software Engineering Conference and Symposium on the Foundations of Software Engineering (ESEC/FSE '21), August 23--28, 2021, Athens, Greece}
\acmISBN{978-1-4503-8562-6/21/08}

\begin{document}

\title{Mono2Micro: A Practical and Effective Tool for Decomposing Monolithic Java Applications to Microservices}

\author{Anup K. Kalia}
\email{anup.kalia@ibm.com}
\author{Jin Xiao}
\email{jinoaix@us.ibm.com}
\affiliation{
  \institution{IBM Research}
  \city{Yorktown Heights}
  \state{NY}
  \country{USA}
  \postcode{10523}
}

\author{Rahul Krishna}
\email{rkrsn@ibm.com}
\author{Saurabh Sinha}
\email{sinhas@us.ibm.com}
\affiliation{
  \institution{IBM Research}
  \city{Yorktown Heights}
  \state{NY}
  \country{USA}
  \postcode{10523}
}

\author{Maja Vukovic}
\email{maja@us.ibm.com}
\affiliation{
  \institution{IBM Research}
  \city{Yorktown Heights}
  \state{NY}
  \country{USA}
  \postcode{10523}
}

\author{Debasish Banerjee}
\email{debasish@us.ibm.com}
\affiliation{
  \institution{IBM Hybrid Cloud, Customer Success}
  \city{Rochester}
  \state{MN}
  \country{USA}
  \postcode{55901}
}

\begin{abstract}
In migrating production workloads to cloud, enterprises often face the daunting task of evolving monolithic applications toward a microservice architecture. At IBM, we developed a tool called to assist with this challenging task. \tool performs spatio-temporal decomposition, leveraging well-defined business use cases and runtime call relations to create functionally cohesive partitioning of application classes. Our preliminary evaluation of showed promising results.

How well does \tool perform against other decomposition techniques, and how do practitioners perceive the tool? This paper describes the technical foundations of \tool and presents results to answer these two questions. To answer the first question, we evaluated \tool against four existing techniques on a set of open-source and proprietary Java applications and using different metrics to assess the quality of decomposition and tool's efficiency. Our results show that \tool significantly outperforms state-of-the-art baselines in specific metrics well-defined for the problem domain. To answer the second question, we conducted a survey of twenty-one practitioners in various industry roles who have used \toolnsp. This study highlights several benefits of the tool, interesting practitioner perceptions, and scope for further improvements.  Overall, these results show that \tool can provide a valuable aid to practitioners in creating functionally cohesive and explainable microservice decompositions.

\end{abstract}



\maketitle


\section{Introduction}
\label{sec:introductions}

Enterprises are increasingly moving their production workloads to cloud to take advantage of cloud capabilities, such as streamlined provisioning of infrastructure and services, elasticity, scalability, reliability, and security. To leverage cloud-native capabilities, monolithic applications have to be decomposed to cloud-native architectures, such as microservices. A \textit{microservice} encapsulates a small and well-defined set of business functionalities and interacts with other services using lightweight mechanisms, often implemented as RESTful APIs~\cite{fowler:2019,newman:2015}. In modernizing legacy applications, enterprises, however, often have to answer the challenging question of \textit{how to transform their monolithic applications to microservices}.

Current strategies for decomposing monolithic applications fall under static- or dynamic-analysis techniques, \ie they typically compute module dependencies using static and/or dynamic analysis and apply clustering or evolutionary algorithms over these dependencies to create module partitions that have desired properties (\eg high cohesion and low coupling). Static approaches \cite{Zhao+CC+2015, Levcovitz+CoRR+2016, Escobar+CLEI+2016, Chen+APSEC+2017, Mazlami+IEEE+2017, Ren+2018+MWA, Taibi+CLOSER+2019, carvalho:2019, Desai+AAAI+2021} suffer imprecision in computing dependencies that is inherent to static analysis. In Java Enterprise Edition (JEE) applications, which are the focus of our work, these techniques face challenges in dealing with dynamic language features, such as reflection, dynamic class loading, context, and dependency injections. In contrast, dynamic techniques (\eg \cite{Patel+ECSMRE+2009, Jin+TSE+2019,Alwis+ICSOC+2018, Fuhr+2011+WCRE}) capture runtime dependencies and thus avoid the imprecision problems. However, a common challenge that still exists for both static and dynamic analysis is computing the alignment of classes and their dependencies with the business functionalities of the application, which is a primary concern in industrial practice.

In this contribution, we show how \toolnsp \cite{Kalia+2020+FSE} based on dynamic analysis achieves the alignment of classes and their dependencies with business functionalities of the application.  \toolnsp \cite{Kalia+2020+FSE}  was developed at IBM and recently in January 2021 made generally available as a product\footnote{https://www.ibm.com/cloud/mono2micro}).

\noindent \textbf{\tool}. We implement a \textit{hierarchical spatio-temporal decomposition} in \tool that dynamically collects runtime traces under the execution of specific business use cases of the application and applies clustering on classes observed in the traces to recommend partitions of the application classes. In this approach, business use cases constitute the \textit{space} dimension, whereas the control flow in the runtime traces expresses the \textit{time} dimension.

\noindent$\bullet$~~\textbf{Business Use Cases}. The space dimension emphasizes the importance of identifying candidate microservices as functionally cohesive groups of classes, each of which implements a small set of well-defined functionalities that can be easily explained to a business user. To implement the space dimension, \tool considers module dependencies specifically in the contexts of business use cases under which they occur. Examples of such business use cases are \subject{Create Account}, \subject{Browse Products}, and \subject{Checkout Products}. In contrast, a technique that analyzes dependencies while ignoring business use cases can recommend partitions that mix different functionalities and, thus, suffer low cohesion. Moreover, the rationale for the computed groupings, agnostic to business use cases, can be hard to explain to a practitioner.

\begin{figure}[t]
  \centering
  \includegraphics[width=1\columnwidth]{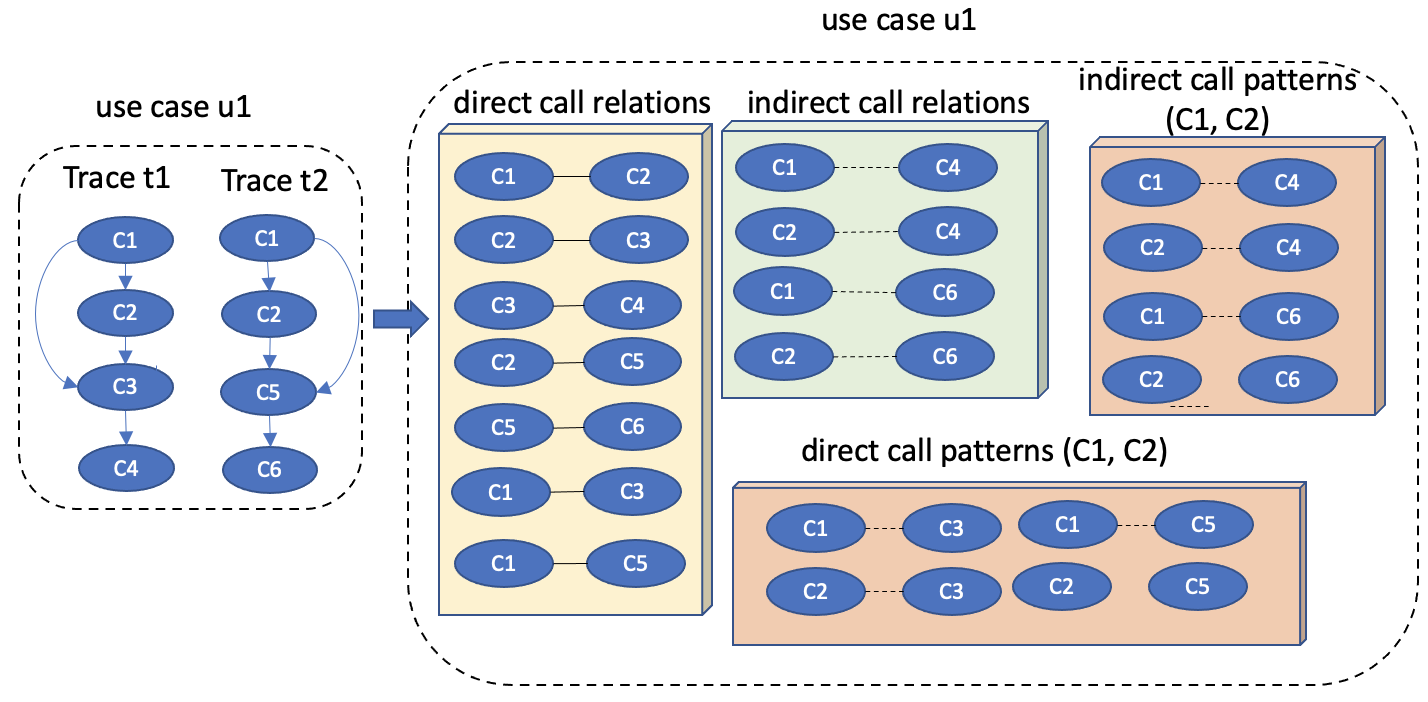}
  \vspace*{-15pt}
  \caption{\small Illustration of execution traces and temporal relations.}
  \label{fig:mono2micro}
 \vspace*{-15pt}
\end{figure}

\noindent$\bullet$~~\textbf{Runtime Call Traces}. The time dimension considers temporal relations and co-occurrence relations among classes extracted from runtime call traces (collected by executing use cases).  Existing techniques in the areas of software repackaging \cite{Patel+ECSMRE+2009,Xiao+2005+CSMR} and microservice extraction
\cite{Alwis+ICSOC+2018,Jin+2018+ICWS,Jin+TSE+2019} analyze direct call relations only. We enhance those approaches in two ways. First, we consider indirect call relations, as shown in Figure~\ref{fig:mono2micro}, that indicate long-range temporal relations among classes. Second, we propose direct call patterns and indirect call patterns to capture the pattern of interaction among classes. The patterns capture the similarity between classes based on how they call other classes through direct or indirect relations across one or more use cases. In Figure~\ref{fig:mono2micro}, $(c_1, c_2)$ and $(c_2, c_3)$ are the examples of direct call relations. $(c_1, c_4)$ and $(c_1, c_6)$ are the examples of indirect call relations. Considering direct call patterns, $c_1$ and $c_2$ are similar based on how they call other classes such as $c_3$ and $c_5$ through direct relations and $c_4$ and $c_6$ through an indirect relations, respectively. We can derive direct and indirect call patterns for other pairs of classes in a similar manner.

\noindent\textbf{Evaluation.}~We describe the technical details of \tool and the results of empirical studies conducted on two sets of JEE applications: four open-source web applications and three proprietary web applications. We evaluate \tool against four well-known baseline approaches from software remodularization and microservices communities i.e., Bunch \cite{mitchell:2006}, FoSCI \cite{Jin+TSE+2019}, CoGCN \cite{Desai+AAAI+2021}, and MEM \cite{Mazlami+IEEE+2017}. We perform the evaluation using five metrics: Inter-Call Percentage (ICP) \cite{Kalia+2020+FSE}, Business Context Purity (BCP) \cite{Kalia+2020+FSE}, Structural Modularity (SM) \cite{Jin+TSE+2019}, Interface Number (IFN) \cite{Jin+TSE+2019} and Non-Extreme Distribution (NED) \cite{Desai+AAAI+2021}. In addition, we conducted a survey among 21 industry practitioners to highlight the importance and benefits of \tool and further scope for improvement.  

Our results indicate that \tool consistently performs well compared with BCP and NED and is competitive with ICP and IFN. Considering SM, \tool did not perform well when compared to Bunch and MEM. However, we observed that high SM scores in such baselines also have higher NED scores indicating extreme distributions. From the survey, we learned several benefits of \tool such as the following. 1) \tool helps implement a Strangler pattern, 2) recommendations generated using \tool capture required business functionalities and are self-explainable, 3) \tool can detect potential unreachable code. In addition, we learned the scope for further improvements of \tool such as the following. 1) minimize the number of changes a user has to make on the top of the recommendations generated, 2) add database interactions and transaction patterns to refine recommendations, and so on.

The rest of the paper is organized as follows. In the next section, we describe the technical details of \tool and illustrate it using an open-source JEE application. Section~\ref{sec:resques} provides the research questions. Section~\ref{sec:evaluation} presents the empirical evaluation. Section~\ref{sec:survey} presents the survey. Section~\ref{sec:summarize} summarizes of research questions. Section~\ref{sec:threats} highlights the threats to the validity of the empirical evaluation and the survey.
Section~\ref{sec:related} discusses related work. Finally, Section~\ref{sec:conclusion} summarizes the paper and lists directions for future research. Section~\ref{sec:ack} provides acknowledgements to everyone who have helped build \toolnsp.

\section{\toolnsp: Technical Approach}
\label{sec:approach}

\begin{figure*}[t]
\vspace*{-10pt}
  \centering
  \includegraphics[width=1\textwidth]{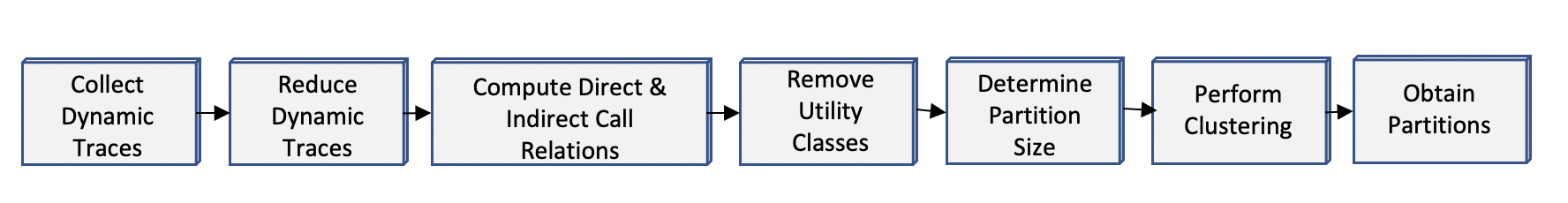}
  \vspace*{-20pt}
  \caption{\small The main steps of the decomposition approach.}
  \vspace*{-10pt}
  \label{fig:process}
\end{figure*}

In this section, we present the technical details of the approach implemented in \toolnsp; Figure~\ref{fig:process} shows the main steps of the approach.  First, we introduce a sample application, \subject{JPetStore}, to illustrate the approach and discuss analysis preliminaries, which consists of trace collection and reduction. Then, we describe the core partitioning technique in the context of the \subject{JPetStore} application.

\subsection{Analysis Preliminaries}
\label{sec:prelims}

\noindent\textbf{Runtime Trace Collection}: Runtime traces are defined as $\mathrm{T}^{(u_i)} = \langle t_1, t_2, \ldots, t_T \rangle$, where each trace $t_i$ is generated by running a use case $u_i$ $\in$ $\mathrm{U}$. A user can manually create such use cases by navigating through the application's user interface (UI) and providing an appropriate label for each use case. If functional test cases are available for an application, one can use them for generating runtime traces. The tests need not be UI test cases, but a test case must correspond to a well-defined application use case (business functionality). Traces record the entries and exits to each function, including the constructors via added probes. For an open-source application \subject{JPetStore} application, we created ten use cases, \eg \textit{update\_item} and \textit{click\_item}. We generated runtime traces by executing the use cases navigating through the UI of the application for each use case. The use cases (traces) cover 37 of the 42 classes (88\% class coverage). The trace obtained via the execution of a use case is a \textit{raw trace}. An example fragment of a raw trace is as follows:

{\footnotesize
\begin{verbatim}
t1,[32],Entering ... PetStoreImpl::getCategory
t2,[32],Entering ... SqlMapCategoryDao::getCategory
t3,[32],Entering ... Category::setCategoryId
t4,[32],Exiting ... Category::setCategoryId
\end{verbatim}
}

Each trace element captures a timestamp, a thread id, and an entry/exit label with a class name and a method signature.

\noindent\textbf{Trace Reduction}: For each use case $u_i$ $\in$ $\mathrm{U}$, we reduce the number of traces in two ways. One, we reduce the total number of traces by considering unique traces. Two, we reduce the length of a trace by removing a redundant sequence of classes that might have invoked due to the presence of a loop. We remove the redundant sequences by converting traces to a representation similar to a calling-context tree (CCT)~\cite{ammons:1997}. Specifically, each trace $t_j \in T^{(u_i)}$ is processed to build a set of CCTs, at the level of class methods, with each tree rooted at an ``entry point'' class that is the first one to be invoked in response to a UI event. Unlike the conventional CCT, in which nodes represent methods~\cite{ammons:1997}, in our CCT, nodes represent classes, thereby further reducing the length of traces.

Below we provide two class-level CCTs that are constructed from the raw traces collected by executing two use cases: \emph{click\_item} and \emph{update\_item}. In the example, \subject{Root} corresponds to an entry-point. For \emph{click\_item}, we obtain one reduced trace (\subject{ViewCatego- ryController} $\rightarrow$ \subject{PetStoreImpl} $\rightarrow$ \subject{SqlMapCategoryDao} $\rightarrow$ \subject{Category}), whereas for \emph{update\_item}, we obtain three reduced traces, each containing two classes; \eg \subject{UpdateCartQuantitiesController} $\rightarrow$ \subject{Cart}.

{\footnotesize
\begin{verbatim}
click\_item, Root, ViewCategoryCont.., PetStoreImpl, SqlMapCategory..
update\_item, Root, UpdateCartQuantitiesController, Cart
update\_item, Root, UpdateCartQuantitiesController, CartIt.
update\_item, Root, UpdateCartQuantitiesController, Item
\end{verbatim}
}

\subsection{Computation of Partitions}

The core of our technique consists of first identifying the similarities among a pair of classes $c_i$ and  $c_j$ where $i$ $\neq$ $j$ and $c_i$ and $c_j$ $\in$ classes $C = \langle c_1, c_2, \ldots, c_C \rangle$ of an application. We identify the similarities by deriving four spatio-temporal different features (1) direct call relations (DCR), (2) indirect call relations (ICR), (3) direct call patterns (DCP), (4) indirect call patterns (ICP). Then, based on the features, we construct a similarity matrix $\mathrm{S}$($c_{k_1}$, $c_{k_2}$) where $c_{k_1}$ and $c_{k_2}$ $\in$ $C$. For the purpose of similarity computation, we consider undirected edges. We apply the hierarchical clustering algorithm on the matrix to decompose the classes into a set of non-overlapping partitions that aims to execute specific business tasks based on business functionalities or use cases.

\noindent\textbf{Direct Call Relations}: A \textit{direct call relation} ($\mathrm{DCR}$) exists between classes $c_{k_1}$ and $c_{k_2}$ if and only if a directed edge ($c_{k_1}$, $c_{k_2}$) exists in an execution trace; \ie a method in $c_{k_1}$ invokes a method in $c_{k_2}$ in a trace. For example, in \subject{JPetStore}, for the \emph{click\_item} use case, the \subject{ViewCategoryController} class calls the \subject{PetStoreImpl} class, whereas for the \emph{update\_item} use case, \subject{UpdateCartQuantitiesContro- ller} calls \subject{Cart}. Thus, \subject{ViewCategoryController} and \subject{PetStoreImpl} have a direct call relation; similarly, \subject{UpdateCartQuantitiesController} and \subject{Cart} also have a direct call relation. 

We leverage the use-case labels $u_i$ associated with an execution trace $t_j$ to compute $\mathrm{DCR}$ as the ratio of the number of use cases where a direct call relation exists between $c_{k_1}$ and $c_{k_2}$, to the union of use cases in which $c_{k_1}$ and $c_{k_2}$ occur: $\mathrm{DCR}$($c_{k_1}$, $c_{k_2}$) = $\frac{|\mathrm{U}_{c_{k_1} \leftrightarrow c_{k_2}}|}{|{\mathrm{U}_{c_{k_1}}} {\cup} {\mathrm{U}_{c_{k_2}}}|}$. For example, if $c_{1}$ participates in two use cases $\{u_{1}, u_{2}\}$, $c_{2}$ participates in three use cases $\{u_{1}, u_{2}, u_{3}\}$, and there is only one direct call relation between them, we compute $\mathrm{DCR}$($c_{1}$, $c_{2}$) as $\frac{1}{3}$. 

\noindent\textbf{Direct Call Pattern}: Based on direct call relations, we derive another spatio-temporal feature \textit{direct call pattern} ($\mathrm{DCP}$) that exists between two classes $c_{k_1}$ and $c_{k_2}$ if and only if there exist an edge $(c_{k_1}, c_{l})$, or $(c_{k_2}, c_{l})$ in the traces; \ie both $c_{k_1}$ and $c_{k_2}$ have a direct call relation with $c_l$ in some execution trace. Whereas $\mathrm{DCR}$ considers the interactions between two classes, $\mathrm{DCP}$ considers whether two classes have a similar pattern of interaction with other classes. We compute $\mathrm{DCP}$($c_{k_1}$, $c_{k_2}$) as follows: $\mathrm{DCP}$($c_{k_1}$, $c_{k_2}$) = 
$\frac{\sum_{c_{} \in C, c_{l} \neq {\{c_{k_1}, c_{k_2}\}}} |\mathrm{U}_{c_{k_1} \leftrightarrow c_{l}} {\cap} \mathrm{U}_{c_{k_2} \leftrightarrow c_{l}}|}{(|C|-2) * (|\mathrm{U}|)}$. To illustrate, consider the call relations for classes $c_1$ and $c_2$ under different use cases shown in Table~\ref{tab:pattern-example}. As shown in Table~\ref{tab:pattern-example},  $c_1$ and $c_2$ do not have a direct call relation. However, $c_1$ and $c_2$ have two direct call relations with $c_{3}$ and $c_{5}$, respectively. We divide the total number of direct call patterns by the total number of possible call patterns $(|C|-2) * (|\mathrm{U}|)$ for $c_1$ and $c_2$ across all use cases. Under the use case $u_{1}$, $c_{1}$ and $c_{2}$ have, in total, two direct call relations with $c_{3}$ and $c_{5}$, respectively.  Therefore, we compute $\mathrm{DCP}$($c_{1}$, $c_{2}$) as $\frac{2}{3*2}$.

\noindent\textbf{Indirect Call Relations}: An \textit{indirect call relation} ($\mathrm{ICR}$) exists between classes $c_{k_1}$ and $c_{k_2}$ if and only if there exists a path ($c_{k_1}$, $c_1$, $\ldots$, $c_p$, $c_{k_2}$), $p \geq 1$, in an execution trace. The indirect call relation ($\mathrm{ICR}$) is calculated as the ratio of the number of use cases where an indirect call relation between $c_{k_1}$ and $c_{k_2}$ occurs to the union of use cases associated with these two classes. For example, in \subject{JPetStore}, for the \emph{browse} use case, the \subject{ViewCategoryController} class has a transitive call relations with the \subject{SqlMapCategoryDao} class and the \subject{Category} class. We calculate $\mathrm{ICR}$($c_{k_1}$, $c_{k_2}$) as $\frac{|\mathrm{U}_{c_{k_1} \Longleftrightarrow c_{k_2}}|}{|{\mathrm{U}_{c_{k_1}}} {\cup} {\mathrm{U}_{c_{k_2}}}|}$.

\begin{table}[t]
\vspace*{-5pt}
  \caption{\small Example to illustrate interaction patterns of two classes ($c_{1}$
    and $c_{2}$) in two use cases ($u_1$ and $u_2$).}
  \label{tab:pattern-example}
  \footnotesize
  \begin{minipage}{.25\linewidth}
  \centering
  \begin{tabular}{ccc}
  \hlineB{2}
     $c_{1}$  & $u_{1}$ & $u_{2}$\\
  \hlineB{1}
    $c_{1}$ & 0 & 0\\
    $c_{2}$ & 0 & 0\\
    $c_{3}$ & 1 & 1\\
    $c_{4}$ & 0 & 0\\
    $c_{5}$ & 1 & 0\\ 
  \hlineB{2}
  \end{tabular}
  \end{minipage}%
  \begin{minipage}{.25\linewidth}
  \centering
  \begin{tabular}{ccc}
  \hlineB{2}
     $c_{2}$   & $u_{1}$ & $u_{2}$\\
  \hlineB{1}
    $c_{1}$ & 0 & 0\\
    $c_{2}$ & 0 & 0\\
    $c_{3}$ & 1 & 0\\
    $c_{4}$ & 0 & 1\\
    $c_{5}$ & 1 & 1\\ 
  \hlineB{2}
  \end{tabular}
  \end{minipage}
  \vspace*{-5pt}
\end{table}

\noindent\textbf{Indirect Call Pattern}: Based on indirect call relations, we define \textit{indirect call pattern} ($\mathrm{ICP}$) as a relation that exists between classes $c_{k_1}$ and $c_{k_2}$ if and only if there exist paths $(c_{k_1}, c_1, \ldots, c_p, c_l)$
and $(c_{k_2}, c_1, \ldots, c_q, c_{l})$, $p \geq 1$ and $q \geq 1$, in the execution traces; \ie both classes have an indirect call relation with a common class $c_l$. $\mathrm{ICP}$ is computed as:
$\mathrm{ICP}$($c_{k_1}$, $c_{k_2}$) = 
$\frac{\sum_{c_{} \in C, c_{l} \neq {\{c_{k_1}, c_{k_2}\}}} |\mathrm{U}_{c_{k_1} \leftrightarrow c_{l}} \cap \mathrm{U}_{c_{k_2} \leftrightarrow c_{l}}|}{(|C|-2) * (|\mathrm{U}|)}$.

\noindent\textbf{Computation of Similarity}: Based on these call relations and patterns, the similarity score between two classes $c_{k_1}$ and $c_{k_2}$ is calculated as: $\mathrm{S}$($c_{k_1}$, $c_{k_2}$) = $\mathrm{DCR}$($c_{k_1}$, $c_{k_2}$) + $\mathrm{DCP}$($c_{k_1}$, $c_{k_2}$) + $\mathrm{ICR}$($c_{k_1}$, $c_{k_2}$) + $\mathrm{ICP}$($c_{k_1}$, $c_{k_2}$). We represent $\mathrm{S}$($c_{k_1}$, $c_{k_2}$) as a similarity matrix.

\noindent\textbf{Hierarchical Clustering}: We use the well-known hierarchical clustering algorithm~\cite{Sibson+1973+CJ} for three reasons. First, it has been investigated in prior work on software modularization \cite{Patel+ECSMRE+2009,Scanniello+ICPC+2010,anquetil:1999} and microservice identification \cite{Jin+TSE+2019}. Second, it has less time complexity compared to the hill-climbing algorithm \cite{mitchell:2006,mahdavi:2003} and genetic algorithms \cite{Jin+TSE+2019,doval:1999,mitchell:2006} (scalability is essential for analyzing large enterprise applications). Third, we assume that monoliths have hierarchical overlapping business processes that need to be separated into microservices and hence a non-parametric approach such as the hierarchical clustering algorithm is appropriate for the setting.

The hierarchical clustering algorithm groups similar objects into clusters (partitions) based on $\mathrm{S}$. The algorithm takes the target number of clusters $n$ as its sole input. Initially, we assign each class $c_k$ $\in$ $C$ to a cluster $P_i$.  During the clustering process, the similarity score $Sim_{i,j}$ between each pair of clusters $i$ and $j$ as $\frac{\sum_{i=0}^{n_{i}}\sum_{n=0}^{n_{j}} S(c_{im}, c_{jn})}{|C_{i}||C_{j}|}$. We merge the pairs with the highest similarity score. We iterate the step until the stopping criterion $n$ is achieved. 

\noindent\textbf{Partitions Explainability}: We obtained five partitions from \subject{JPet- Store} using $n=5$.  We provide the details of the partitions and corresponding use cases in Section~\ref{sec:datasets}.

We observe the five partitions represent five different microservices, respectively: 1) \emph{init}, 2) \emph{item}, 3) \emph{register}, 4) \emph{order}, and 5) \emph{browse}. Each microservice is represented as a group of classes where each class has a mapping to a tuple of use cases. For example, in case of the \emph{init} microservice, \subject{ListOrdersController} and \subject{ViewOrderControl- ler} are mapped with the $\langle$init$\rangle$ tuple whereas \subject{SearchProductsContro- ller} is mapped with the $\langle$init, search$\rangle$ tuple. The mapping of a class with a tuple indicates that a class is invoked under one or more use cases present in the tuple. Based on overlapping use cases across tuples, we find classes under the \emph{init} microservice are aligned with the \emph{init} specific business functionality. Similarly in case of the \emph{register} microservice, \subject{SignonController} are mapped with the $\langle$init, login\_user$\rangle$ tuple whereas \subject{AccountValidator} is mapped with the $\langle$ register\_user, submit\_user $\rangle$ tuple. Here, both the tuples may not have overlapping use cases, however, semantically both the tuples are related to the \emph{register} microservice. Thus, we observe classes under the \emph{register} microservice are aligned with the \emph{register} specific business functionality.  

Accordingly, the collection of tuples of use cases for each partition provides the explainability for the partition in terms of the business functionalities for users to comprehend the partitions' correctness.

\section{Empirical Evaluation}
\label{sec:evaluation}

For the evaluation, we followed this general procedure: (1) we collected execution traces based on use cases, (2) we generated reduced paths using CCTs, and (3) we ran the implementation of our partitioning approach to generate partitions.

\subsection{Subject Applications}

We used seven JEE applications for the evaluation, consisting of four open-source applications and three proprietary enterprise applications. Table~\ref{tab:subjects} presents the data about the applications and use-case-based execution traces collected on the applications. We chose open-source applications for the following reasons. First, all of them are JEE web applications that are available as deployable and runnable applications. Second, they have a monolithic architecture. Third, they have been used in prior evaluations in academic research~\cite{Jin+TSE+2019,Desai+AAAI+2021}. The open-source applications are small and have class-coverage rates ranging from 66\% to 88\%. The proprietary applications are larger but have lower coverage rates, in particular for \subject{App2}, and \subject{App3}.

\begin{table}[t]
\vspace*{-5pt}
\footnotesize
\centering
\caption{\small Subject applications and use-case-based traces used in the evaluation.}
\label{tab:subjects}
\begin{tabular}{p{1.3cm}p{0.5cm}p{0.8cm}p{0.2cm}p{1.5cm}p{2cm}}
\hlineB{2}
Apps & Classes & Methods & \#UC & Class Coverage & Method Coverage\\
\hlineB{1}
DayTrader \footnote{https://github.com/WASdev/sample.daytrader7} & 109 & 969 & 83 &  73 (66\%) & 428 (44\%)\\
AcmeAir \footnote{https://github.com/acmeair/acmeair} & 33 & 163 & 11 & 28 (84\%) & 108 (66\%) \\
JPetStore \footnote{https://github.com/KimJongSung/jPetStore} & 66 & 350 & 44 & 36 (54\%) & 236 (67\%)\\
Plants \footnote{https://github.com/WASdev/sample.mono-to-ms.pbw-monolith} & 37 & 463 & 43 & 25 (67\%) & 264 (57\%)\\
\hlineB{1}
App1 & 82 & 449 & 15 & 50 (60\%) & 247 (55\%) \\
App2 & 245 & 333 & 7  & 60 (24\%)  & 280 (8\%)\\ 
App3 & 1286 & 12,066 & 21 & 241 (19\%) & 1517 (12\%)\\
\hlineB{2}
\end{tabular}
\vspace*{-5pt}
\end{table}

\subsection{Baseline Techniques}

We compare \tool with four baselines: FoSCI \cite{Jin+TSE+2019}, CoGCN \cite{Desai+AAAI+2021}, Bunch \cite{mitchell:2006} and MEM \cite{Mazlami+IEEE+2017}. We selected them based on the following criteria. 1) their source code is available to replicate their methods; 2) they are well-known techniques from microservice identification (FoSCI, CoGCN, and MEM) and software re-modularization (Bunch) research areas; and 3) they require minimal manual data preparation for usage. There are other relevant baselines such as ServiceCutter~\cite{Gysel+2016+ESOCC} that requires significant manual effort in generating the inputs such as the entity-relationship model (ERM) from an application. We realized that such effort is intractable and cannot be scaled to applications with more than 1000 classes.

\begin{itemize}[wide=0pt]
  
\item \textbf{FoSCI}\footnote{https://github.com/wj86/FoSCI/releases} \cite{Jin+TSE+2019}, creates functional atoms using a hierarchical clustering approach and then merges the atoms using a genetic algorithm to compute partition recommendations. For FoSCI, we considered both structural and conceptual connectivity.

\item \textbf{CoGCN}\footnote{https://github.com/utkd/cogcn}\cite{Desai+AAAI+2021} proposes an approach to partition a monolith applications by minimizing the effect of outlier classes that might be present in the embeddings of other classes. For CoGCN, we construct their three matrices: EP($i$, $p$), C($i$, $j$), and In($i$, $j$). EP($i$, $p$) suggests if a class $i$ is present in an entry point $p$, C($i$, $j$) suggests if two classes $i$ and $j$ are present in an entry point, and In($i$, $j$) suggests if $i$ and $j$ related by the inheritance relationship.

\item \textbf{Bunch}\footnote{https://github.com/ArchitectingSoftware/Bunch} \cite{mitchell:2006} needs an external module dependency graph (MDG) as its input to generate partitions. For Bunch, we consider a version of its hill-climbing algorithm. We considered the nearest-ascend hill climbing (NAHC) as suggested by Saeidi {\etal} \cite{saeidi:2015}.

\item \textbf{MEM} \cite{Mazlami+IEEE+2017}\footnote{https://github.com/gmazlami/microserviceExtraction-backend} considers the minimum spanning tree (MST) approach that uses Kruskal's algorithm \cite{Kruskal+AMS+1956} for computing the minimum spanning trees. We consider their logical and semantic coupling strategies to generate partitions.
\end{itemize}

Based on the input data obtained from the subject applications using \tool, we created data converters to convert the input data to the format required by each of these four baselines.

\subsection{Metrics}
We provide five metrics to measure the effectiveness of partitions recommended using \toolnsp.

\begin{itemize}[wide=0pt]
\item \textbf{SM} \cite{Jin+TSE+2019} measures the modularity quality of partitions as the structural cohesiveness of classes within a partition ($m_i$) (scoh) and coupling (scop) between the partitions ($M$). It is computed as $\frac{1}{M}$$\sum^{M}_{i=1}$$scoh_i-$ $\frac{1}{(M(M-1))/2}$$\sum^{M}_{i{\ne}j}$$scop_{i,j}$. $scoh_i$ is computed as $\frac{\mu_i}{m^2_i}$ where $\mu_i$ refers the number of edges within a partition $m_i$ and $scop_{i,j}$ is computed as $\frac{\sigma_{i,j}}{2*(m_i*m_j)}$ where $\sigma_{i,j}$ refers the number of edges between partitions $m_i$ and $m_j$. Higher the value of SM, better is the recommendation.

\item \textbf{ICP} \cite{Kalia+2020+FSE} measures the percentage of runtime calls occurring between two partitions $icp_{i, j} = c_{i,j} / \sum_{i=1, j=1, i \neq j}^{M} c_{i,j}$, where $c_{i,j}$ represents the number of call between partition $i$ and partition $j$. Lower the value of ICP, better is the recommendation.

\item \textbf{BCP} \cite{Kalia+2020+FSE} measures the average entropy\footnote{https://docs.scipy.org/doc/scipy/reference/generated/scipy.stats.entropy.html} of business use cases per partition. The use cases for a partition consists of all use-case labels associated with its member classes. A partition is considered functionally cohesive if it implements a small set of use cases. BCP is computed as $\frac{1}{M}$$\sum_{i=1}^{M}$$bcp_{i}$, where $bcp_{i}$ is computed as -$\sum_{j=1}^{m_i}$ $\frac{1}{|m_{i}|}$ $log$ ($\frac{1}{|m_{i}|}$) where $\frac{1}{|m_{i}|}$ is a vector of the size $m_{i}$ where $m_{i}$ represents the number of use cases for a partition $i$ $\in$ $M$. For example, given a partition with 3 use cases, $\frac{1}{|m_{i}|}$ is represented as [1/3, 1/3, 1/3]. Lower values for BCP indicate better recommendations.

\item \textbf{IFN} \cite{Jin+TSE+2019} measures the number of interfaces in a microservice. IFN is computed as $\frac{1}{N}$$\sum_{i=1}^{N}$$ifn_{i}$ where $ifn_{i}$ is the number of interfaces in a microservice where $N$ is the total number of microservices. Lower values of IFN indicates better recommendations.

\item \textbf{NED} \cite{Desai+AAAI+2021} measures how evenly the size of a microservice is. It is measured as 1 - $\frac{\sum_{i=1, k not extreme}^{K}{n_k}}{|N|}$ where k ranges in \{5, 20\} \cite{scanniello:2010a}. Lower values of NED indicates better recommendations. NED was originally proposed by Wu {\etal} \cite{Wu+ICSME+2005} to evaluate the extremity of a microservice distribution.
\end{itemize}

\subsection{Hyperparameter Settings}


For hyperparameter settings, we first consider the number of partitions to consider for each approach. Several approaches have been used in prior work for determining a partition size. Some of these require users to choose a cut-point \cite{Xiao+2005+CSMR,Patel+ECSMRE+2009}: \ie a value between 0 and the maximum height of a dendogram obtained using the hierarchical-clustering algorithm. Other approaches provide a stopping criteria; \eg Jin {\etal} \cite{Jin+TSE+2019} use Jaccard distance values greater than three to merge clusters. Such approaches require users to determine a value for each application, which in practice the user may not always know. For our experiments, we chose to adopt the approach suggested by Scanniello {\etal} \cite{Scanniello+ICPC+2010} where we take a range of cluster sizes (partition size values) starting from $\frac{N}{2}$, N > 0 and keep going downward to a value greater than 1 where $N$ represents the number of classes. Here, for small applications ($N \leq 50$), we use a slower rate ($N-2^{i}$), whereas for larger applications ($N \geq 100$), we consider $\frac{N}{2^{i}}$ where $i \geq 1$. The strategy is applicable to \toolnsp, CoGCN, FoSCI, and MEM but not Bunch. Bunch does not provide an explicit option to provide a partition size as its input, rather it provides three agglomerative output options to generate partitions: top level, median level, and the detailed level. For FoSCI, we consider the \emph{diff} as 3 for all the applications except for App$_2$ where the number of functional atoms flattened when \emph{diff}=1. For other hyperparameters for FoSCI, Bunch, CoGCN, and MEM, we consider the values provided by the authors for each approach.

\section{Results}
\label{sec:resques}

We compare \tool against four baselines using five evaluation metrics. We also conducted a survey of \tool with industry practitioners to get their feedback.  In particular, our evaluation and the survey aims to address the following research questions:

\begin{description}
\item[RQ1:] How does \tool perform based on the quality of partitioning using a set of metrics?

\item[RQ2:] How fast is \tool's partitioning?

\item[RQ3:] How helpful do industry practitioners find \tool in refactoring their monolithic applications? 
\end{description}

\subsection{Partitioning Quality (RQ1)}

Table~\ref{tab:evaluations-open}~and~\ref{tab:evaluations-web} present the comparison of \tool with FoSCI, CoGCN, Bunch and MEM for seven applications across five metrics. For each application, we created a range of partitions and obtained the score for all the metrics. We removed the outliers and computed median scores for each metric. For Bunch, considering only three partition values, the IFN score for \subject{Daytrader} and the SM score for \subject{Jpetstore} got omitted once we remove the outliers. Table~\ref{tab1-heatmap-A} indicates the overall winners across all approaches.


Considering BCP and NED, \tool significantly outperformed other approaches as shown in Table~\ref{tab1-heatmap-A}. \tool winning in terms of BCP indicates that use-case-based partitions are more functionally cohesive. \tool winning for NED implies that the majority of the partitions generated by \tool contain 5 to 20 classes. The result is due to the non-parametric approach based on hierarchical clustering rather than multi-objective optimization and parametric methods like k-means that other baselines use. We observed that for App3, \tool lost to FoSCI in NED, indicating the possible adjustments for the NED constraints for larger applications.

Considering ICP and IFN,  \tool performed better than other approaches. However, the performance does not hold across the majority of applications. In terms of ICP,  \tool outperformed other approaches for \subject{Daytrader}, \subject{Jpetstore}, and \subject{App2}. Followed by \tool, FoSCI performed better than other approaches for \subject{App2} and \subject{App3} and CoGCN for \subject{Acmeair}. For \subject{App2}, using FoSCI, we obtained a significantly lower ICP score, whereas the NED score obtained is significantly higher compared to the approaches. This suggests that high non-extreme distribution values might have led to monolithic partitions, thereby lowering the ICP scores. However, this did not hold for \subject{Acmeair} and \subject{App3} where CoGCN and FoSCI performed well for both ICP and NED, respectively.

For SM, Bunch outperformed all other approaches for \subject{Daytrader}, \subject{Acmeair}, \subject{App1}, \subject{App2}, and \subject{App3} followed by MEM that outperformed other approaches for \subject{Jpetstore}, \subject{Plants}. Bunch internally uses a function that optimizes for cohesion and coupling based on internal and external edges, respectively. We assume that this might be the reason for high SM values for Bunch. Although, we observed that in the majority of the applications, the NED scores for Bunch are higher than other approaches that suggest non-extreme distribution. The result is due to Bunch's technique that might lead to obtaining large monolithic partitions at the cost of high SM.

\def\cca#1{\cellcolor{blue!#10}\color{black}{#1}} 

\begin{table}[!t]
\caption{\small Heatmap showing the overall winners among all the approaches (0 indicates the lowest score whereas 6 indicates the highest score). Here M2M is \toolnsp.}
\label{tab1-heatmap-A}
\centering
\small
\begin{tabular}{|l|l|l|l|l|l|}
\hlineB{2}
 & \textbf{M2M} & \textbf{FoSCI} & \textbf{CoGCN} & \textbf{Bunch} & \textbf{MEM}\\
\hlineB{1}
\textbf{BCP} & \cca{5} &  \cca{0} & \cca{1} & \cca{1} & \cca{0}\\
\textbf{ICP} & \cca{3} & \cca{2} & \cca{1} & \cca{0} &	\cca{1}\\
\textbf{SM} & \cca{0} & \cca{0} & \cca{0} & \cca{5} &	\cca{2}\\
\textbf{IFN} & \cca{2} & \cca{1} & \cca{1} & \cca{1} &	\cca{1}\\
\textbf{NED} & \cca{5} & \cca{2} & \cca{1} & \cca{0} &	\cca{0}\\
\hlineB{2}
\end{tabular}
\end{table}


\begin{table}[!htb]
\centering
\small
\caption{\small Evaluation results for all the open source applications in terms of median BCP, ICP, SM, IFN and NED scores obtained for a range of partitions. Here M2M is \toolnsp.}
\label{tab:evaluations-open}
\begin{tabular}{|l|r|r|r|r|r|}
\hlineB{2}
\rowcolor{gray!12}\textbf{Daytrader} & \textbf{M2M} & \textbf{FoSCI} & \textbf{CoGCN} & \textbf{Bunch} & \textbf{MEM} \\
\hlineB{2}
\textbf{BCP} & \cellcolor{blue!10}{0.907} & 1.641 & 1.073 & 1.858 & 1.965 \\
\textbf{ICP} & \cellcolor{blue!10}{0.346} & 0.748 & 0.455 & 0.572 & 0.355 \\
\textbf{SM} & 0.078 & 0.092 & 0.086 & \cellcolor{blue!10}{0.269} & 0.089 \\
\textbf{IFN} & \cellcolor{blue!10}{1.922} & 3.489 & 2.880 & $-$ & 4.200 \\ 
\textbf{NED} & \cellcolor{blue!10}{0.338} & 0.697 & 0.663 & 0.582 & 1.000\\
\hlineB{2}\multicolumn{1}{c}{}& \multicolumn{1}{c}{}& \multicolumn{1}{c}{}&\multicolumn{1}{c}{} &\multicolumn{1}{c}{} &\multicolumn{1}{c}{} \\[-0.9em]\hlineB{2}

\rowcolor{gray!12}\textbf{AcmeAir} & \textbf{M2M} & \textbf{FoSCI} & \textbf{CoGCN} & \textbf{Bunch} & \textbf{MEM} \\
\hlineB{1}
\textbf{BCP} & \cellcolor{blue!10}{0.953} & 1.539 & 1.221 & 1.545 & 1.827\\
\textbf{ICP} & 0.527 & 0.706 & \cellcolor{blue!10}{0.444} & 0.55 & 0.589\\
\textbf{SM} & 0.072 & 0.095 & 0.038 & \cellcolor{blue!10}{0.177} & 0.097\\
\textbf{IFN} & 3.375 & 4.375 & \cellcolor{blue!10}{2.846} & 3.875 & 4.333\\
\textbf{NED} & 0.429 & 0.407 & \cellcolor{blue!10}{0.250} & 0.692 & 0.464\\
\hlineB{2}\multicolumn{1}{c}{}& \multicolumn{1}{c}{}& \multicolumn{1}{c}{}&\multicolumn{1}{c}{} &\multicolumn{1}{c}{} &\multicolumn{1}{c}{} \\[-0.9em]\hlineB{2}
\rowcolor{gray!12}\textbf{Jpetstore} & \textbf{M2M} & \textbf{FoSCI} & \textbf{CoGCN} & \textbf{Bunch} & \textbf{MEM} \\
\hlineB{1}
\textbf{BCP} & \cellcolor{blue!10}{1.625} & 2.181 & 1.905 & 2.433 & 2.496\\
\textbf{ICP} & \cellcolor{blue!10}{0.333} & 0.478 & 0.582 & 0.477 & 0.434\\
\textbf{SM} & 0.054 & 0.044 & 0.091 & $-$ & \cellcolor{blue!10}{0.124}\\
\textbf{IFN} & \cellcolor{blue!10}{1.857} & 3.750 & 2.533 & 7.948 & 3.429\\
\textbf{NED} & \cellcolor{blue!10}{0.257} & 0.516 & 0.392 & 0.667 & 1.000\\
\hlineB{2}\multicolumn{1}{c}{}& \multicolumn{1}{c}{}& \multicolumn{1}{c}{}&\multicolumn{1}{c}{} &\multicolumn{1}{c}{} &\multicolumn{1}{c}{} \\[-0.9em]\hlineB{2}
\rowcolor{gray!12}\textbf{Plants} & \textbf{M2M} & \textbf{FoSCI} & \textbf{CoGCN} & \textbf{Bunch} & \textbf{MEM} \\
\hlineB{1}
\textbf{BCP} & \cellcolor{blue!10}{1.690} & 2.593 & 2.338 & 2.902 & 1.902\\
\textbf{ICP} & 0.381 & 0.682 & 0.571 & 0.501 & \cellcolor{blue!10}{0.320}\\
\textbf{SM} & 0.078 & 0.135 & 0.133 & 0.155 & \cellcolor{blue!10}{0.210}\\
\textbf{IFN} & 6.000 & 4.875 & 4.875 & 6.357 & \cellcolor{blue!10}{4.750}\\
\textbf{NED} & \cellcolor{blue!10}{0.038} & 0.538 & 0.500 & 0.346 & 0.231\\
 \hlineB{2}
\end{tabular}
\end{table}

\begin{table}[!htb]
\centering
\small
\caption{\small Evaluation results for all the web enterprise applications in terms of median BCP, ICP, SM, IFN and NED scores obtained for a range of partitions. Here M2M is \toolnsp.}
\label{tab:evaluations-web}
\begin{tabular}{|l|l|l|l|l|l|}
\hlineB{2}
\rowcolor{gray!12}\textbf{App1} & \textbf{M2M} & \textbf{FoSCI} & \textbf{CoGCN} & \textbf{Bunch} & \textbf{MEM} \\
\hlineB{2}
\textbf{BCP} & \cellcolor{blue!10}{0.888} & 1.433 & 1.347 & 1.209 & 1.429\\
\textbf{ICP} & \cellcolor{blue!10}{0.214} & 0.58 & 0.456 & 0.426 & 0.489\\
\textbf{SM} & 0.184 & 0.143 & 0.061 & \cellcolor{blue!10}{0.281} & 0.216\\
\textbf{IFN} & \cellcolor{blue!10}{2.750} & 5.100 & 3.923 & 2.933 & 5.400\\
\textbf{NED} & \cellcolor{blue!10}{0.438} & \cellcolor{blue!10}{0.438} & 0.471 & 0.565 & 1.000\\
\hlineB{2}\multicolumn{1}{c}{}& \multicolumn{1}{c}{}& \multicolumn{1}{c}{}&\multicolumn{1}{c}{} &\multicolumn{1}{c}{} &\multicolumn{1}{c}{} \\[-0.9em]\hlineB{2}
\rowcolor{gray!12}\textbf{App2} & \textbf{M2M} & \textbf{FoSCI} & \textbf{CoGCN} & \textbf{Bunch} & \textbf{MEM} \\
\hlineB{2}
\textbf{BCP} & 0.404 & 0.828 & 0.424 & \cellcolor{blue!10}{0.297} & 0.543\\
\textbf{ICP} & 0.329 & \cellcolor{blue!10}{0.021} & 0.759 & 0.267 & 0.561\\
\textbf{SM} & 0.137 & 0.119 & 0.060 & \cellcolor{blue!10}{0.238} & 0.137\\
\textbf{IFN} & 1.625 & \cellcolor{blue!10}{0.333} & 3.050 & 1.786 & 3.607\\
\textbf{NED} & \cellcolor{blue!10}{0.121} & 0.690 & 0.262 & 0.500 & 1.000\\
\hlineB{2}\multicolumn{1}{c}{}& \multicolumn{1}{c}{}& \multicolumn{1}{c}{}&\multicolumn{1}{c}{} &\multicolumn{1}{c}{} &\multicolumn{1}{c}{} \\[-0.9em]\hlineB{2}
\rowcolor{gray!12}\textbf{App3} & \textbf{M2M} & \textbf{FoSCI} & \textbf{CoGCN} & \textbf{Bunch} & \textbf{MEM} \\
\hlineB{2}
\textbf{BCP} & 1.511 & 1.495 & \cellcolor{blue!10}{1.275} & 1.432 & 1.542\\
\textbf{ICP} & 0.626 & \cellcolor{blue!10}{0.233} & 0.899 & 0.647 & 0.758\\
\textbf{SM} & 0.133 & 0.045 & 0.029 & \cellcolor{blue!10}{0.270} & 0.09\\
\textbf{IFN} & 11.548 & 39.000 & 17.500 & \cellcolor{blue!10}{6.921} & 7.879\\
\textbf{NED} & 0.921 & \cellcolor{blue!10}{0.890} & 1.000 & 0.934 & 1.000\\
 \hlineB{2}
\end{tabular}
\end{table}

\subsection{Runtime (RQ2)}
Table~\ref{tab1-time} shows the median time in seconds taken by each approach to generate partitions. We compared the approaches to find that \tool takes significantly less time than FoSCI, CoGCN, and MEM to generate partitions. Bunch with the hill-climbing approach takes the least amount of time. In addition, we find that FoSCI with a genetic algorithm takes the most amount of time followed by CoGCN that takes a neural network approach, and MEM, which takes the minimum spanning tree approach.

\begin{table}[!t]
\caption{\small The median time in seconds required to generate partitions for all the applications using different approaches. Here M2M is \toolnsp.}
\label{tab1-time}
\centering
\small
\begin{tabular}{|l|l|l|l|l|l|}
\hlineB{2}
\rowcolor{gray!12}\textbf{Apps} & \textbf{M2M} & \textbf{FoSCI} & \textbf{CoGCN} & \textbf{Bunch} & \textbf{MEM}\\
\hlineB{2}
\textbf{Daytrader} & 0.604 & 2601 & 23.06 & 0.045 & 1.680 \\
\textbf{AcmeAir} & 0.049 & 181.4 & 10.92 & 0.044 & 1.620 \\
\textbf{Jpetstore} & 0.103 & 210.9 & 14.41 & 0.030 & 1.600 \\
\textbf{Plants} & 0.049 & 87.24 & 11.01 & 0.042 & 3.500 \\
\textbf{App1} & 0.133 & 718.3 & 17.04 & 0.040 & 1.634 \\
\textbf{App2} & 0.185 & 6.292 & 19.83 & 0.043 & 1.660 \\
\textbf{App3} & 5.005 & 6886 & 38.45 & 0.062 & 2.570\\
\hlineB{2}
\end{tabular}
\end{table}

\subsection{User Survey (RQ3)}
\label{sec:survey}

We surveyed industry practitioners to understand how they perceive \tool. For the survey, we created a questionnaire with 20 questions adopted from existing surveys \cite{Zhang+ICSA+2019, Fritzsch+ICSME+2019, Kirby+ICPC+2021, Wang+EMSE+2021}. First, we conducted a pilot study with 4 participants to refine the questionnaire and estimate the total time required to complete the survey. Next, we sent out the survey questionnaire to 32 participants who have tried \tool. Among the 32 participants, 21 participants returned the survey results. The participants belonged to the following job roles: 1) technical sales (21.1\%), 2) software architect (21.1\%), 3) software developer (15.8\%), and 4) others. Considering software industry experience, 1) 84.2\% participants have 10+ years of experience, 2) 10.5\% participants with 5-10 years of experience, and 3) the rest with 1-3 years. For microservices development, 1) 36.8\% participants have more than five years of experience, 2) 26.3\% participants with three years of experience, 3) 15.8\% participants with one year of experience, 4) 10.5\%  participants with two years of experience, and 5) 10.5\% participants with four years of experience. We asked the participants questions as shown in Table~\ref{tab1-survey-qs}. We provide their response below.

\begin{table*}[!htb]
\caption{Survey questions categorized into four groups.}
\label{tab1-survey-qs}
\footnotesize
\centering
\resizebox{\linewidth}{!}{
\begin{tabular}{|C{1.6cm}|p{12cm}|C{1.2cm}|C{1cm}|C{0.7cm}|}
\hlineB{2}
\rowcolor{gray!12}\multicolumn{1}{V{2}c|}{\textbf{Category}} & \multicolumn{1}{|c|}{\textbf{Questions}} & \multicolumn{1}{|C{1.2cm}|}{\textbf{Response Format}} & \multicolumn{1}{|c|}{\textbf{Free Text?}} & \multicolumn{1}{|cV{2}}{\textbf{Responses}}~ \bigstrut\\
\hlineB{2}
\multicolumn{1}{c}{~}&\multicolumn{1}{c}{~}&\multicolumn{1}{c}{~}&\multicolumn{1}{c}{~}&\multicolumn{1}{c}{~}\\[-0.9em]\hlineB{1}
\multirow{5}{*}{\parbox{2cm}{{\textbf{Preliminary Information}}}} & Q1. How long have you used \tool? & MCQ & $\times$ & \protect\Cref{fig:q_1}\bigstrut\\
& Q2. How many applications have you analyzed with \tool? & MCQ & $\times$ & \protect\Cref{fig:q_2}  \bigstrut\\
& Q3. What kinds of applications have you analyzed with \tool & MCQ & $\times$ & \protect\Cref{fig:q_3}\bigstrut\\
& Q4. What kind of supporting resources for \tool have you used? & MCQ & $\times$ & \protect\Cref{fig:q_4}\bigstrut\\
& Q5. Have you used applications other than \tool? If so, which one? & open & \checkmark & \protect\Cref{fig:q_5} \bigstrut\\\hlineB{1}
\multicolumn{1}{c}{~}&\multicolumn{1}{c}{~}&\multicolumn{1}{c}{~}&\multicolumn{1}{c}{~}&\multicolumn{1}{c}{~}\\[-0.9em]\hlineB{1}
\multirow{5}{*}{\parbox{1.8cm}{\textbf{Architectural Decisions}}} & Q6. Is \tool helpful in implementing the Strangler pattern in decomposing apps? & 1 to 5 scale& $\times$ & \protect\Cref{fig:q_6} \bigstrut\\
& Q7. Is \tool helpful in implementing a Domain-Driven Design (DDD) Pattern in decomposing apps? &1 to 5 scale& $\times$ & \protect\Cref{fig:q_7} \bigstrut\\
& Q8. Do the partitions created by \tool are aligned with business functionality of the apps? &1 to 5 scale& $\times$ & \protect\Cref{fig:q_8} \bigstrut\\
& Q9. Which of the following relationship types are most important to decomposing your app? a. Structural b. Semantic c. Evolutionary &1 to 5 scale& $\times$ &  $\cdot$ \bigstrut\\
& Q10. Which of the following factors \tool should consider apart from runtime traces and use cases? a. static call graphs, b. class name similarity, c. change history, d. commit similarity, e. contributor similarity, f. database interaction patterns, g. database transactions  &1 to 5 scale & $\times$ & $\cdot$ \bigstrut\\
\hlineB{1}
\multicolumn{1}{c}{~}&\multicolumn{1}{c}{~}&\multicolumn{1}{c}{~}&\multicolumn{1}{c}{~}&\multicolumn{1}{c}{~}\\[-0.9em]\hlineB{1}
\multirow{5}{*}{\parbox{1.8cm}{\textbf{Running \tool}}} & Q11. The partitions generated by \tool support independent or mutually exclusive business functionalities? & 1 to 5 scale & $\times$ & \protect\Cref{fig:q_11} \bigstrut\\
& Q12. If dependencies exist in business functionalities across partitions, what are the main reasons? & open & \checkmark & $\cdot$ \bigstrut\\
& Q13. Did \tool's recommendation provides new perspective to your application that you find useful? & open & \checkmark & $\cdot$ \bigstrut\\
& Q14. When collecting traces did you execute the application use cases manually or did you have an automated suite of functional test cases? & MCQ &$\times$ & \protect\Cref{fig:q_14} \bigstrut\\
& Q15. Which experiences with use cases and \tool matches your experience? & MCQ & $\times$ & \protect\Cref{fig:q_15} \bigstrut\\
\hlineB{1}
\multicolumn{1}{c}{~}&\multicolumn{1}{c}{~}&\multicolumn{1}{c}{~}&\multicolumn{1}{c}{~}&\multicolumn{1}{c}{~}\\[-0.9em]\hlineB{1}
\multirow{5}{*}{\parbox{2cm}{\textbf{Explanability, Configuration, Performance}}} & Q16. Did you find the ``explainability'' of partitions, as indicated by use-case labels, to be valuable? & 1 to 5 scale& $\times$ & \protect\Cref{fig:q_16} \bigstrut\\
& Q17. How many and what kind of changes did you make to the original partition suggestions? & MCQ & $\times$ & \protect\Cref{fig:q_17}\bigstrut\\
& Q18. What kinds of changes did you make? & MCQ & $\times$ & \protect\Cref{fig:q_18}\bigstrut\\
& Q19. When you chose the number of partitions what was more valuable to you? Why? & MCQ & $\times$ & \protect\Cref{fig:q_19}\bigstrut\\
& Q20. Is \tool is fast enough to generate recommendations that it does not slow down my workflow? & 1 to 5 scale & $\times$ & \protect\Cref{fig:q_20}\bigstrut\\
\hlineB{1}
\end{tabular}}
\end{table*}

\subsubsection{Preliminary Information}

In terms of preliminary information, we observed the following. 1) Based on Q1, most participants (42.1\%) have used \tool for 1-2 months. 2) Based on Q2, we observed most participants used it on a few applications to date, e.g., 36.8\% on 2-3 applications and 31.6\% of the participants on one application. 3) Based on Q3, we find 35.8\% of the participants have used \tool for sample applications. The responses to Q1, Q2, and Q3, respectively, are expected since \tool was recently made generally available (GA) in early 2021. 4) Based on Q4, most participants (33.3\%) relied on \tool's user guide. 5) Based on Q5, the majority of the participants mentioned they did not use any tools for refactoring before \tool. One participant mentioned about CAST \footnote{https://www.castsoftware.com/products/code-analysis-tools} and ADDI \footnote{https://www.ibm.com/products/app-discovery-and-delivery-intelligence} whereas another participant mentioned about the Transformation Advisor tool \footnote{https://www.ibm.com/garage/method/practices/learn/ibm-transformation-advisor}. We find the response interesting, considering there are plenty of refactoring tools available from academia. The tool availability aspect deserves further study. Our current hypothesis is that it is important to have active product support to gain popularity. Users are looking for tools that can support their modernization methodologies, as we will discuss shortly.

\begin{result}
\textbf{Lesson 1}: Enterprise users are inclined toward using supported industry tools.
\end{result}
\begin{figure*}[t]
\centering
    \subfloat[\footnotesize{\tool usage time (Q1)}]{
        \includegraphics[width=0.22\linewidth]{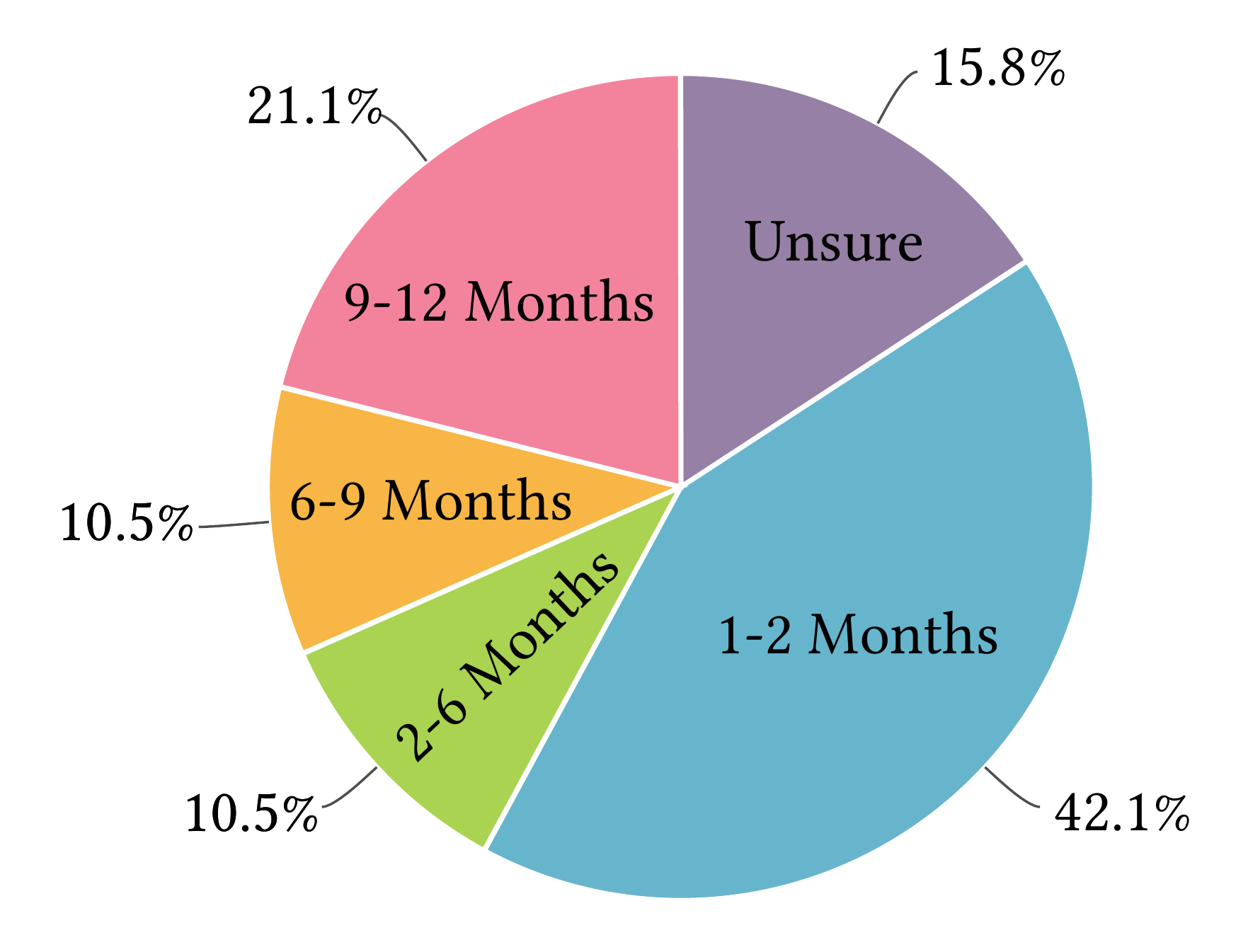}
        \label{fig:q_1}
    }%
    \subfloat[\footnotesize{No. applications assessed (Q2)}]{
        \includegraphics[width=0.22\linewidth]{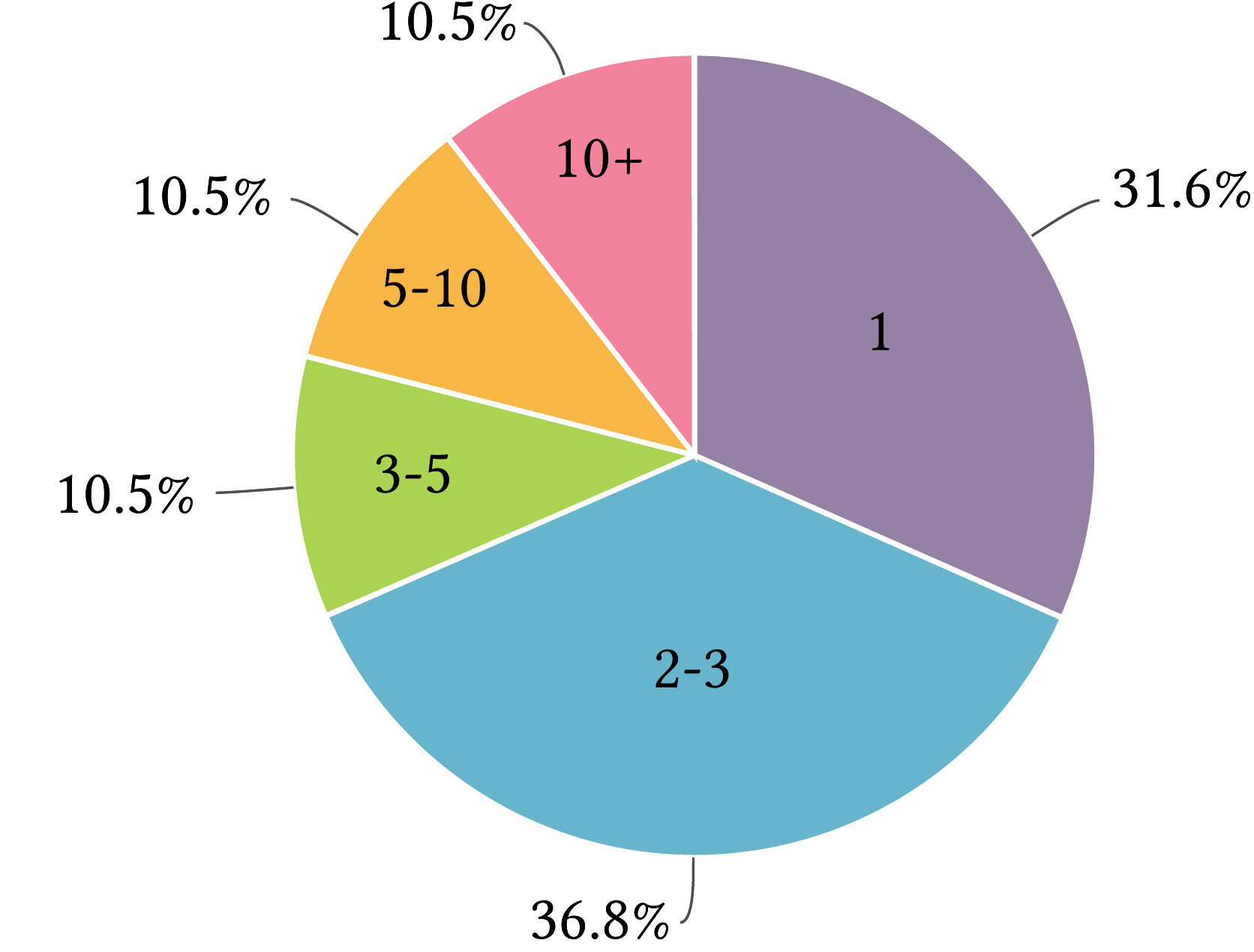}
        \label{fig:q_2}
    }%
    \subfloat[\footnotesize{Assessed application types (Q3)}]{
        \includegraphics[width=0.22\linewidth]{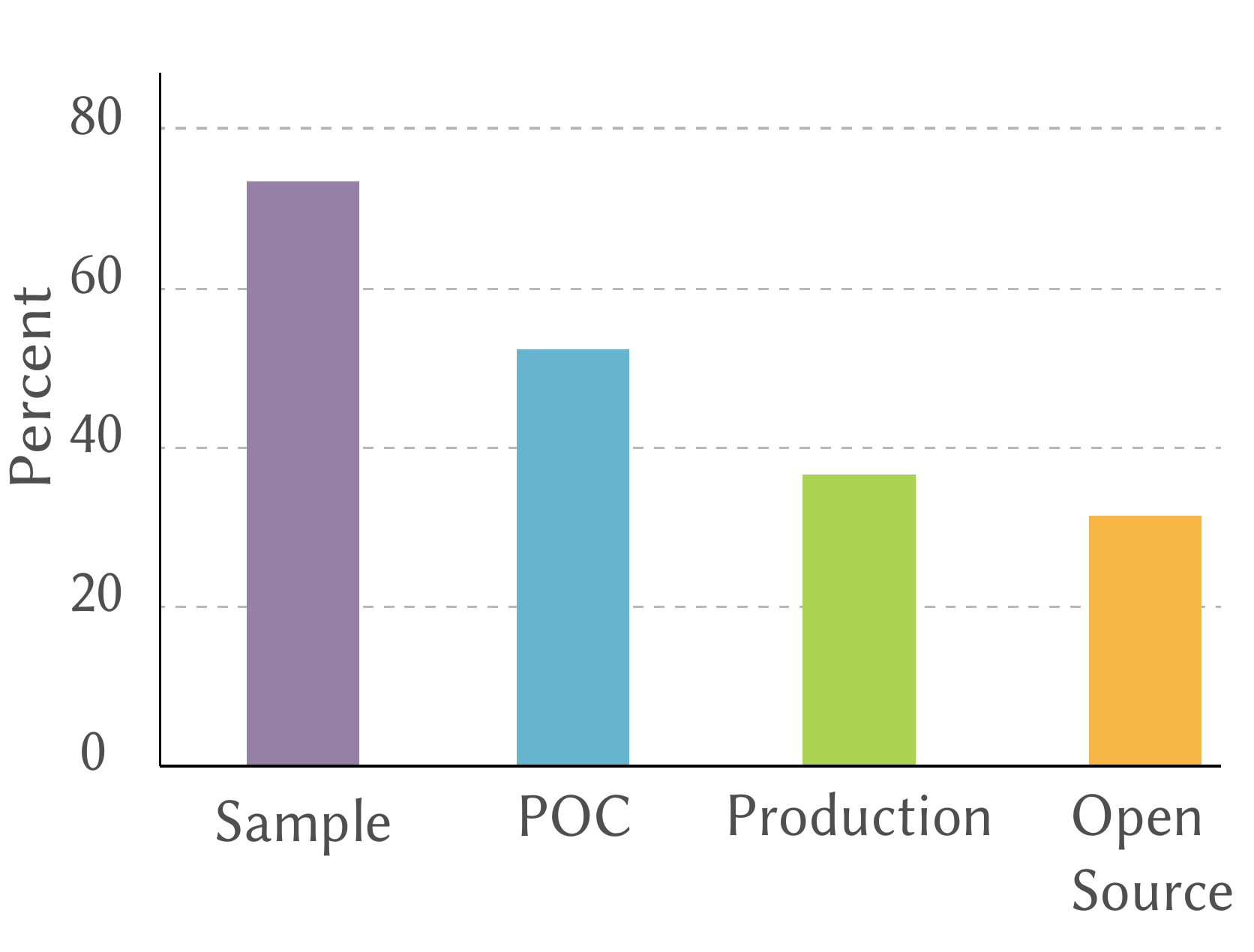}
        \label{fig:q_3}
    }%
    \subfloat[\footnotesize{Supporting resource used (Q4)}]{
        \includegraphics[width=0.22\linewidth]{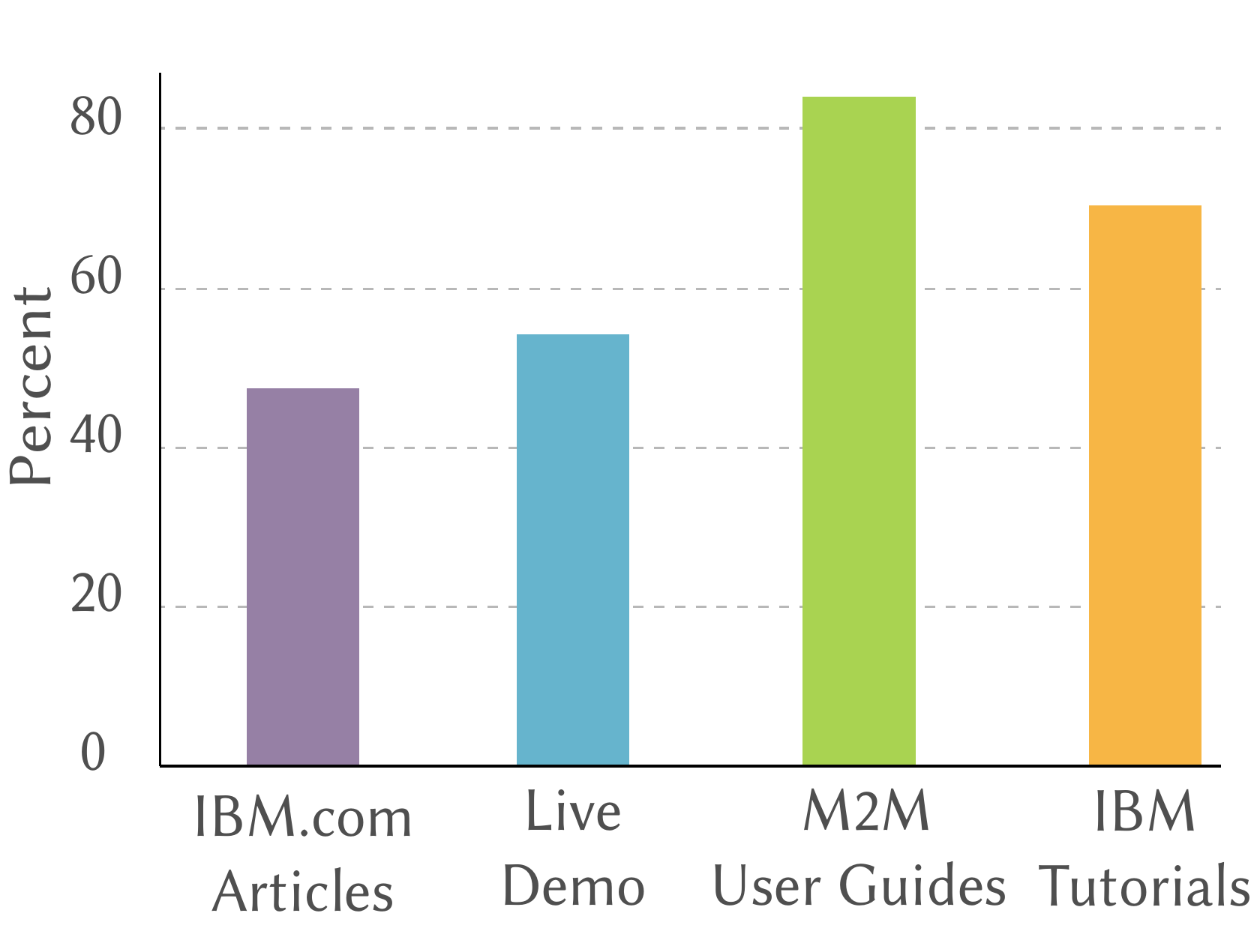}
        \label{fig:q_4}
    }%
    \\
    \subfloat[\footnotesize{\tool alternatives  used (Q5)}]{
        \includegraphics[width=0.22\linewidth]{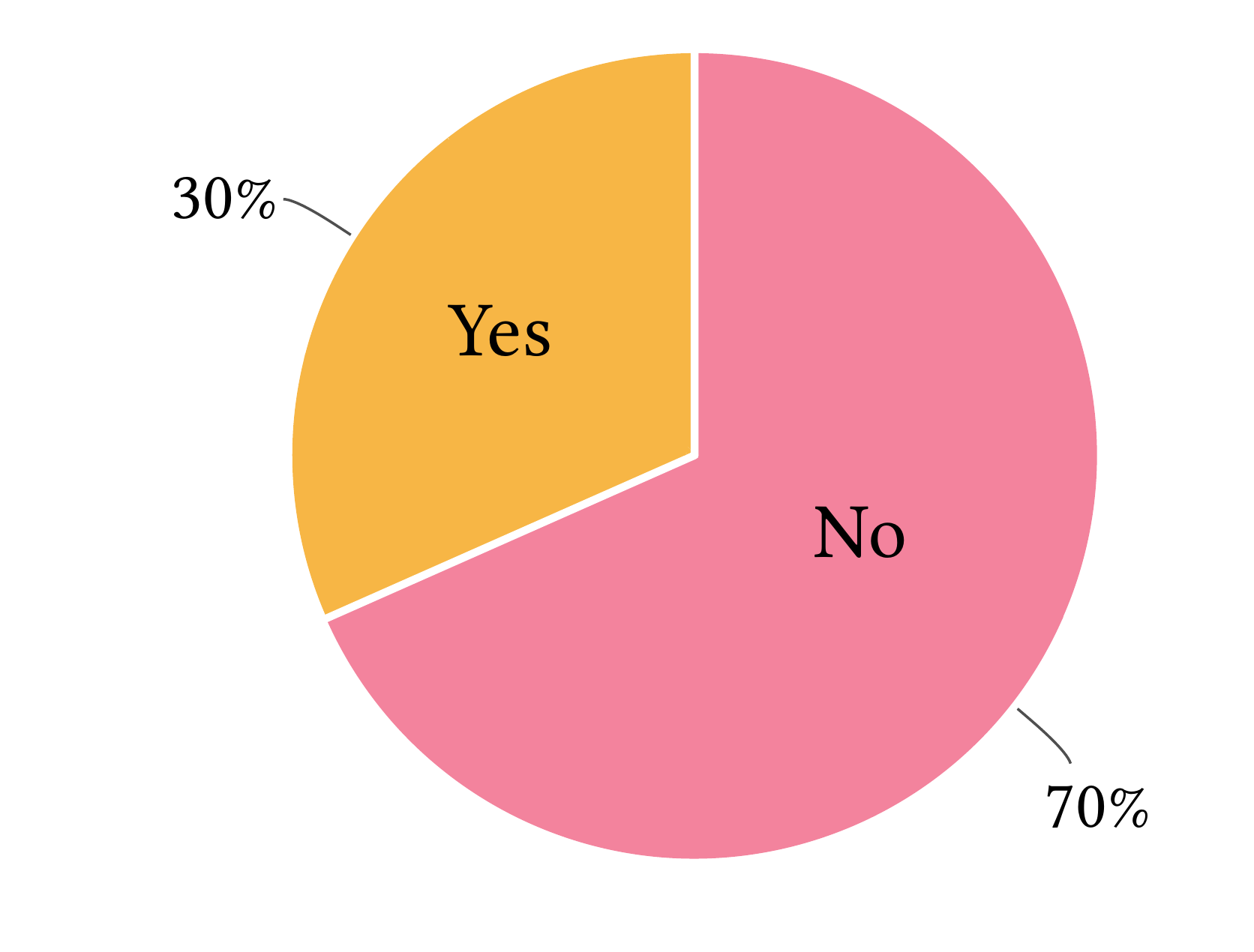}
        \label{fig:q_5}
    }%
    \subfloat[\footnotesize{Strangler pattern observed (Q6)}]{
        \includegraphics[width=0.22\linewidth]{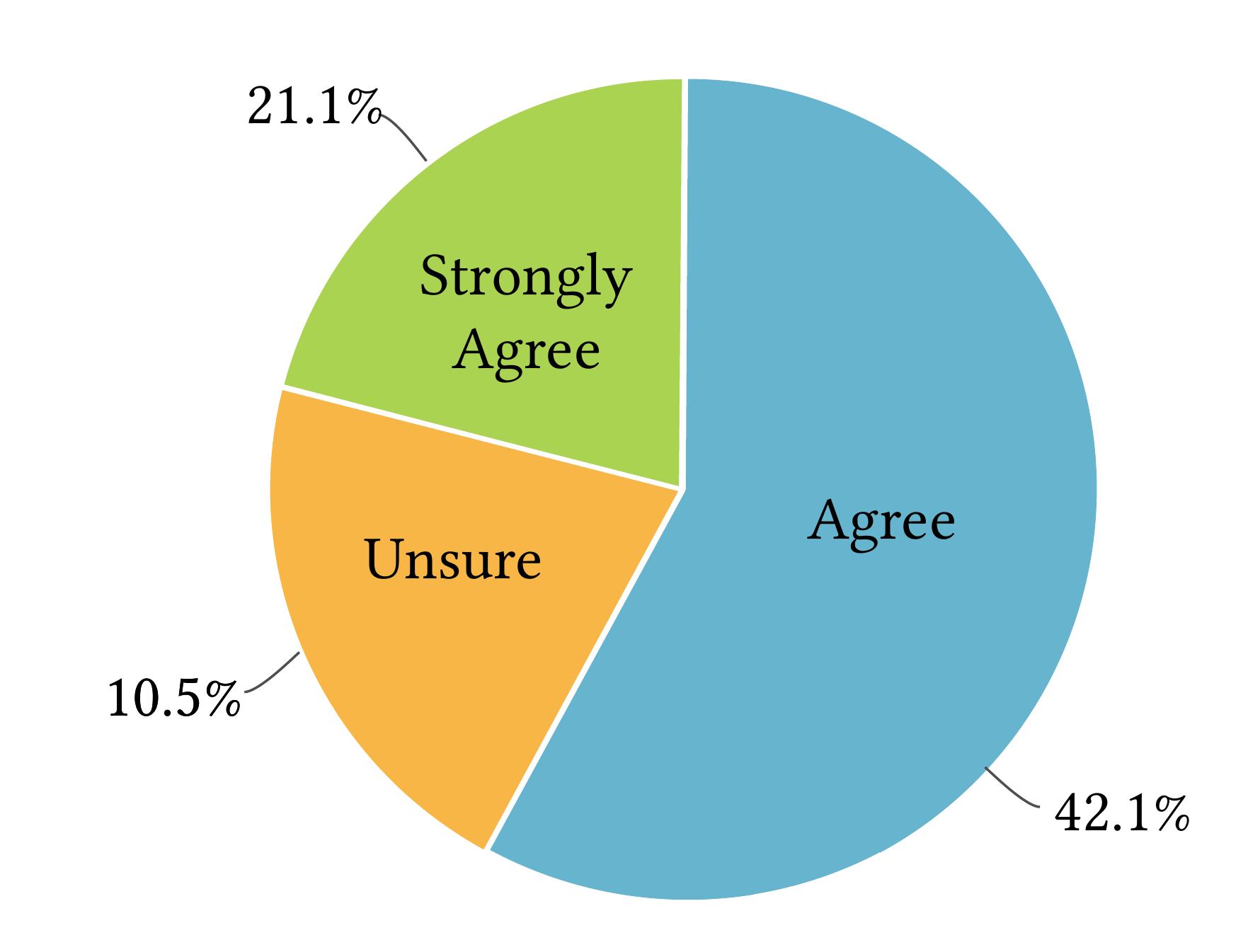}
        \label{fig:q_6}
    }%
    \subfloat[\footnotesize{DDD pattern observed (Q7)}]{
        \includegraphics[width=0.22\linewidth]{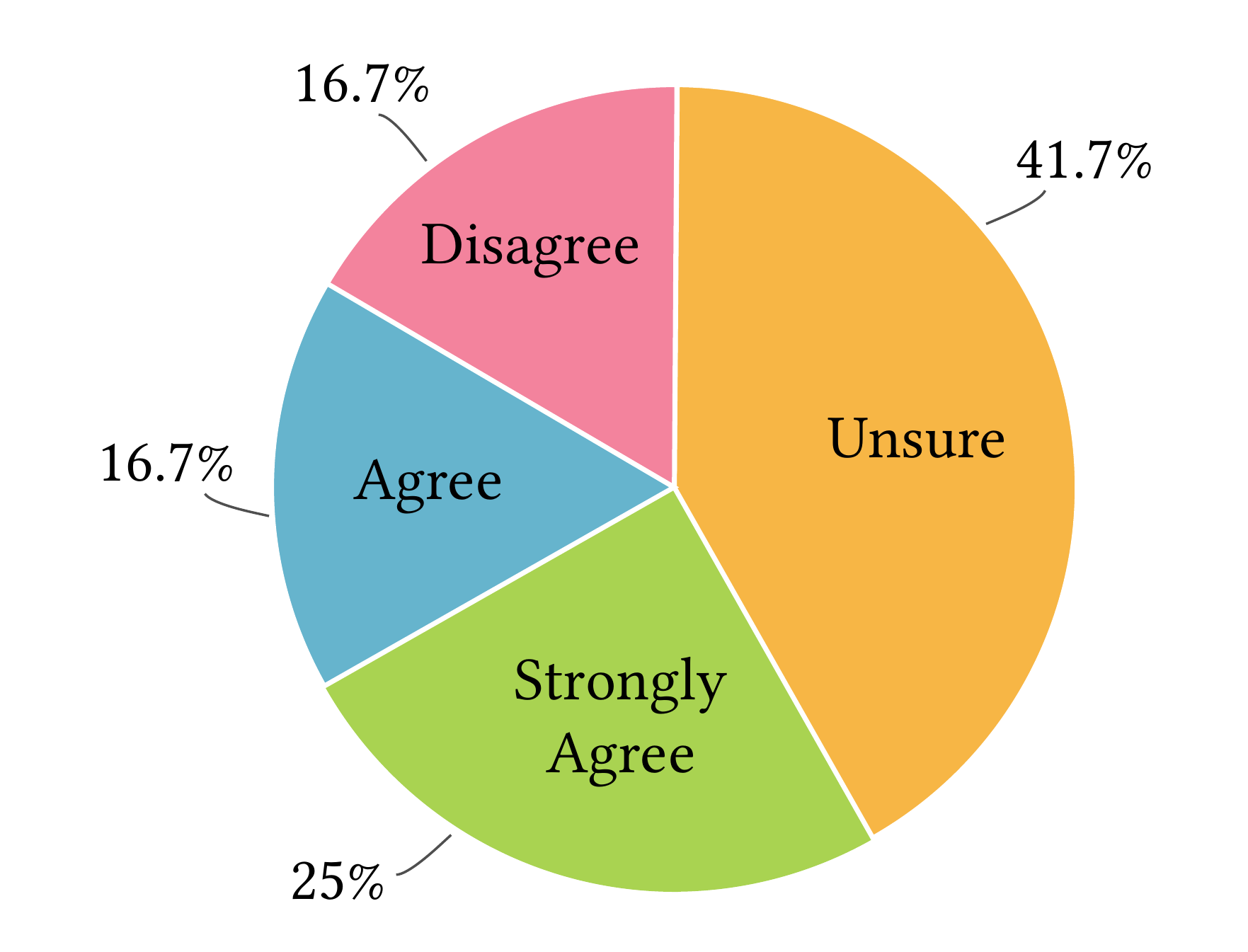}
        \label{fig:q_7}
    }%
    \subfloat[\footnotesize{\tool partitions align with business functions (Q8)}]{
        \includegraphics[width=0.22\linewidth]{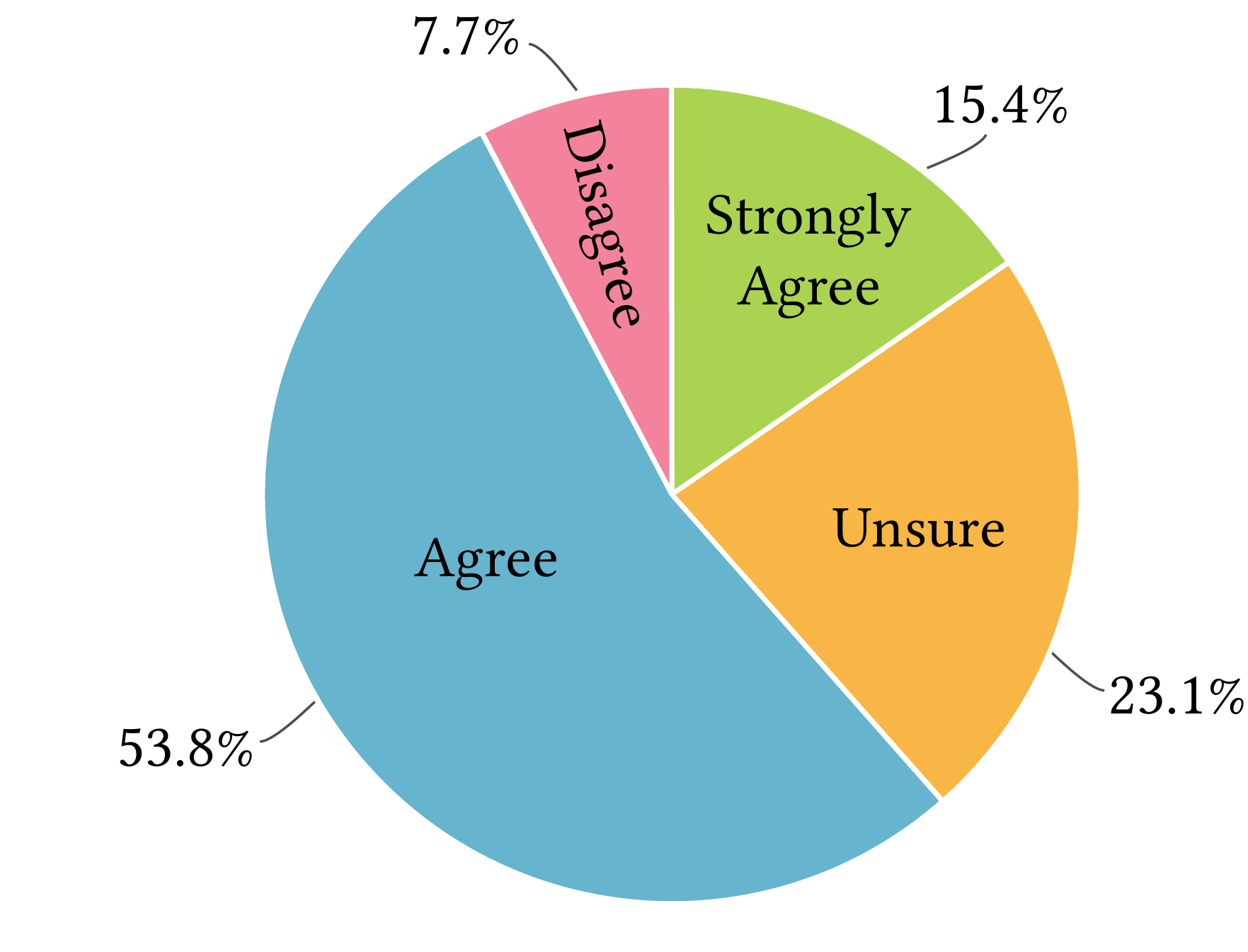}
        \label{fig:q_8}
    }%
    \\
    \subfloat[\footnotesize{Partitions were independent and/or mutually exclusive of one another (Q11)}]{
        \includegraphics[width=0.22\linewidth]{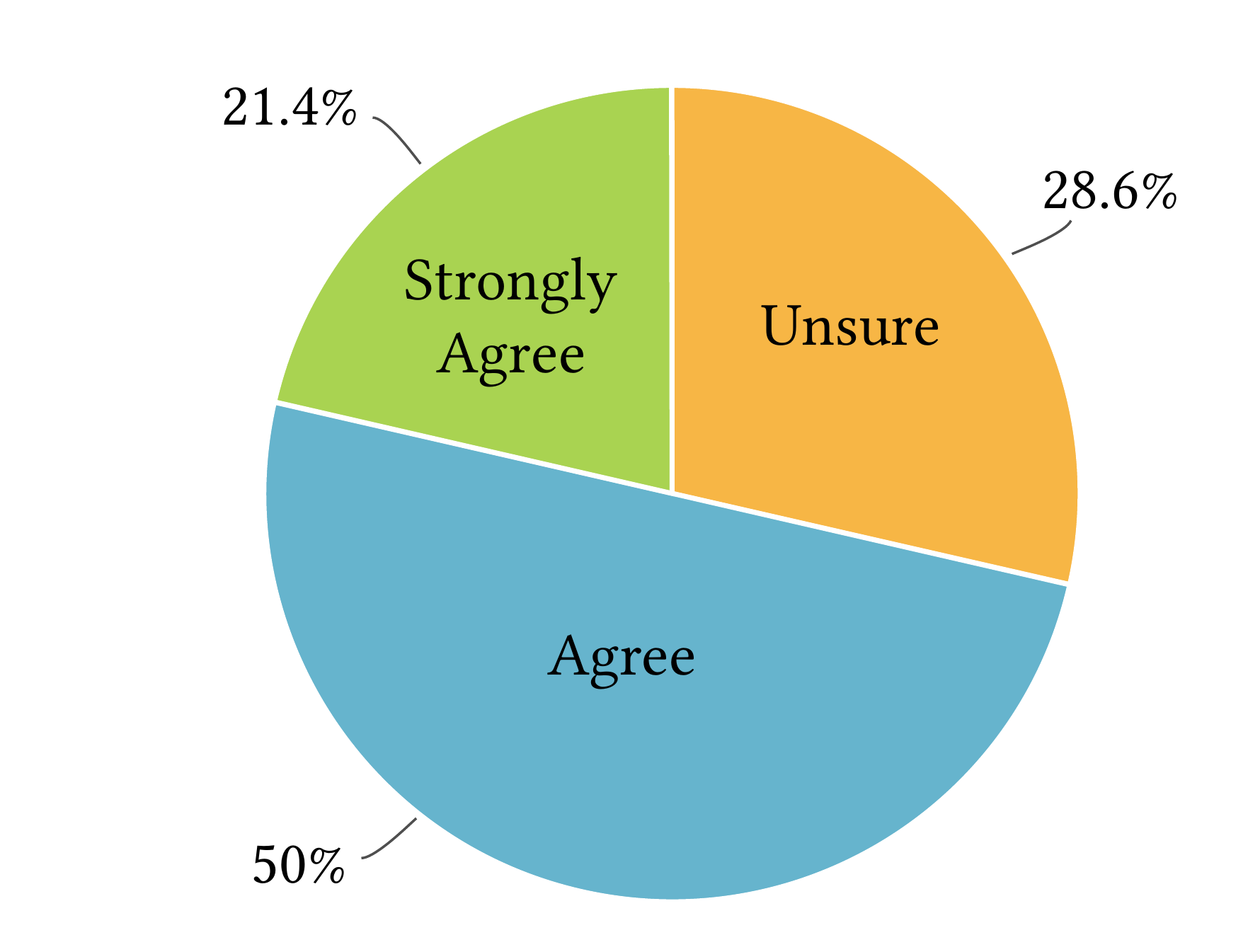}
        \label{fig:q_11}
    }%
    \subfloat[\footnotesize{Trace collection strategy (Q14)}]{
        \includegraphics[width=0.22\linewidth]{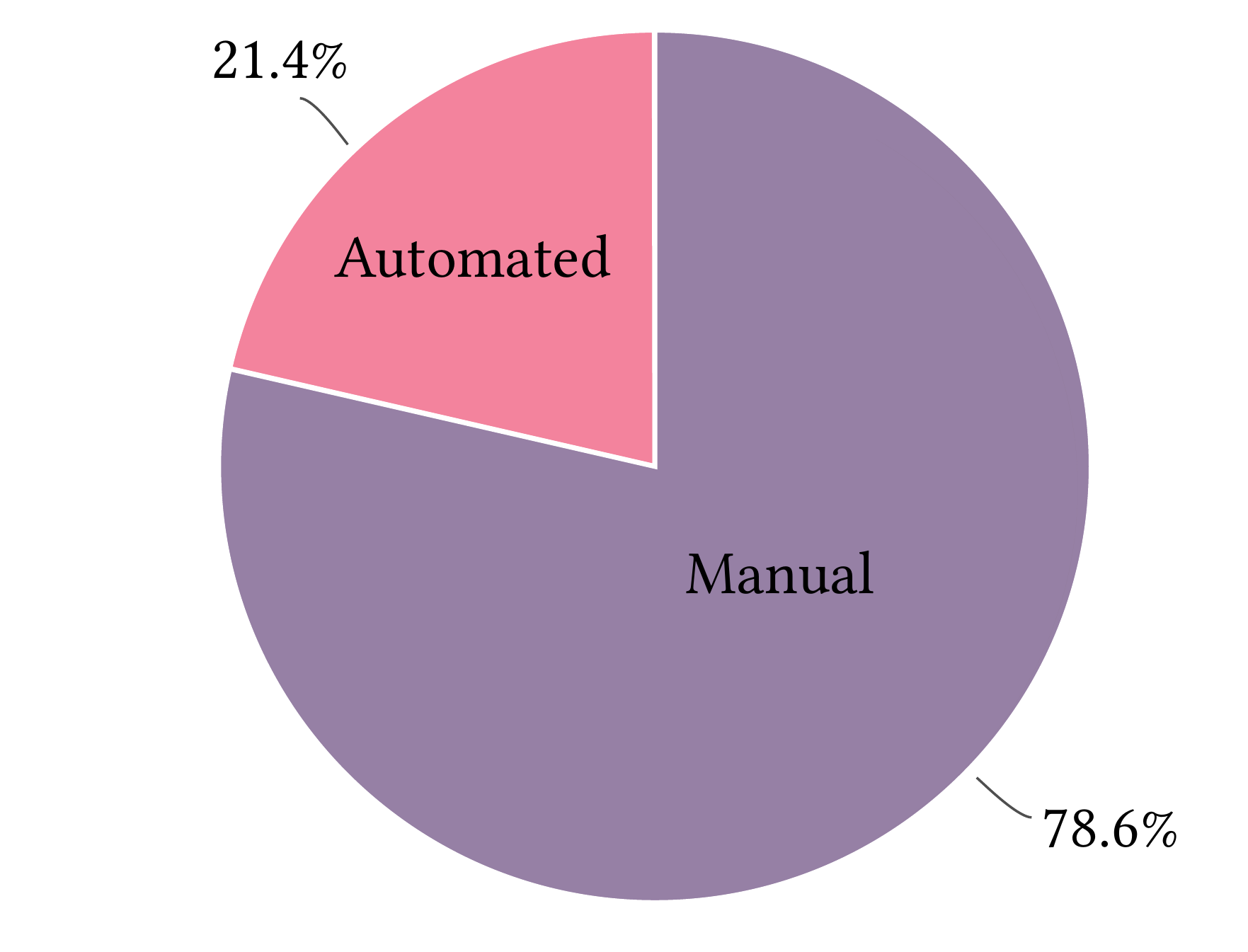}
        \label{fig:q_14}
    }%
    \subfloat[\footnotesize{Experiences with use cases and \tool (Q15)}]{
        \includegraphics[width=0.22\linewidth]{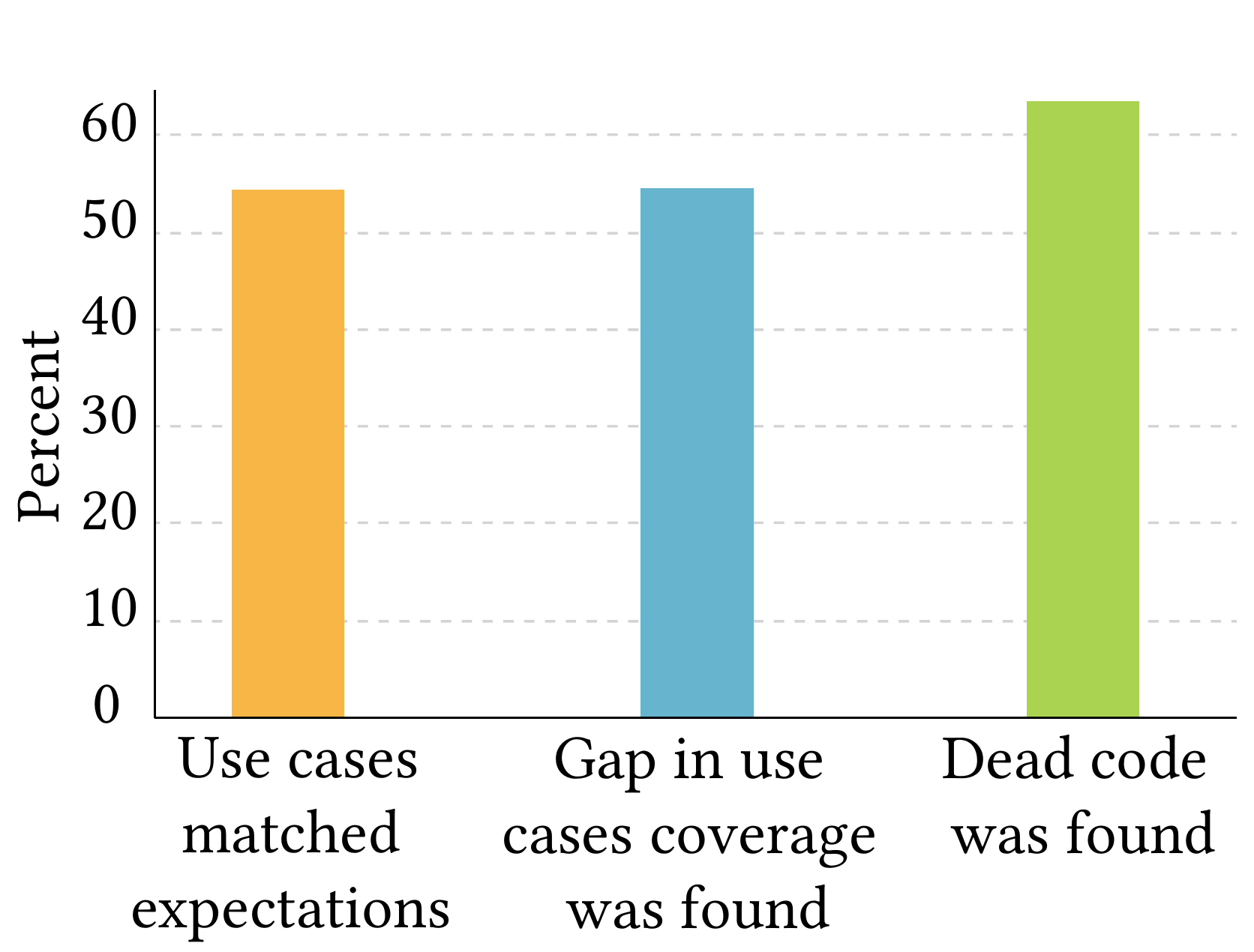}
        \label{fig:q_15}
    }%
    \subfloat[\footnotesize{Value of use-case labels (Q16)}]{
        \includegraphics[width=0.22\linewidth]{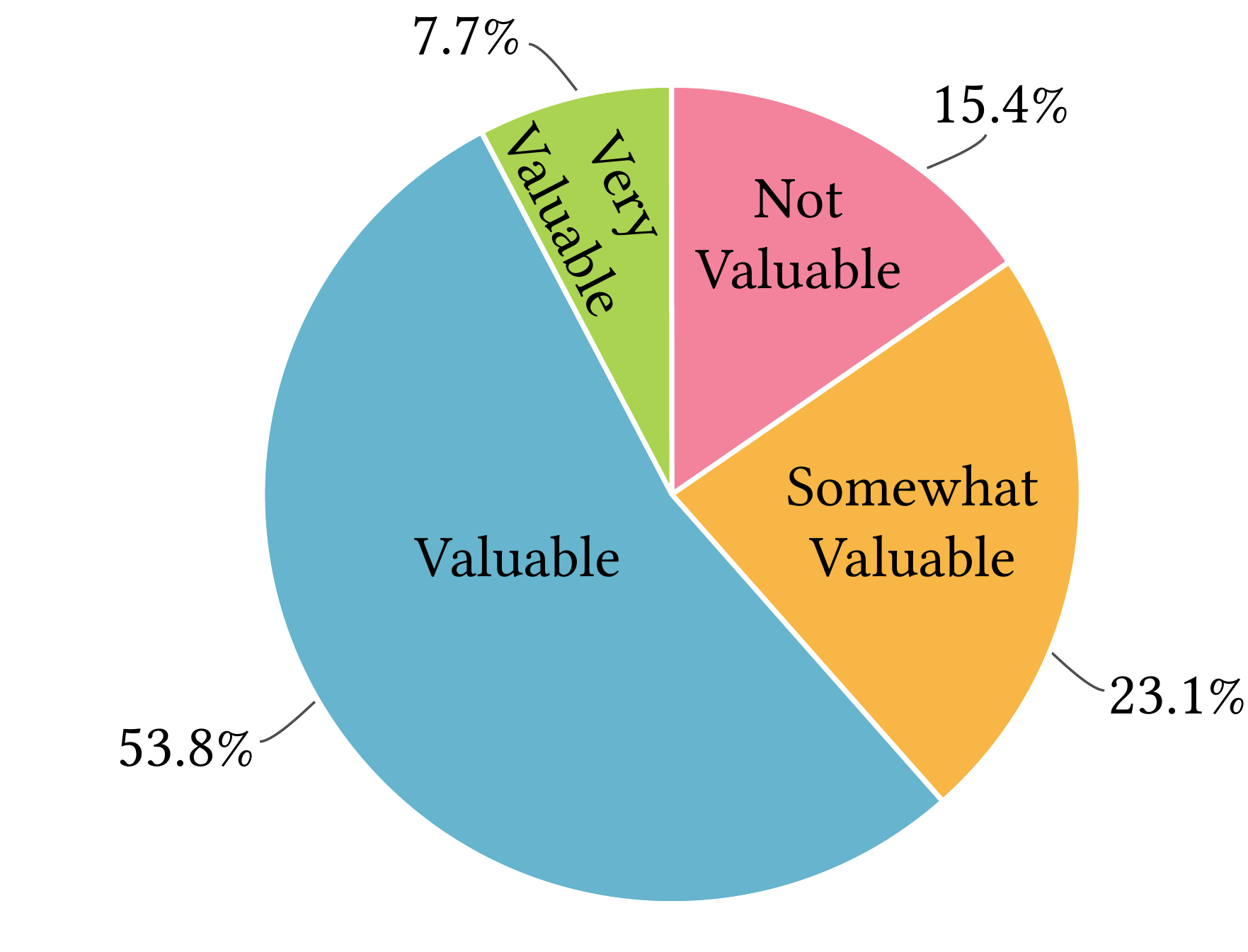}
        \label{fig:q_16}
    }%
    \\
    \subfloat[Changes to partitions from \tool (Q17)]{
        \includegraphics[width=0.22\linewidth]{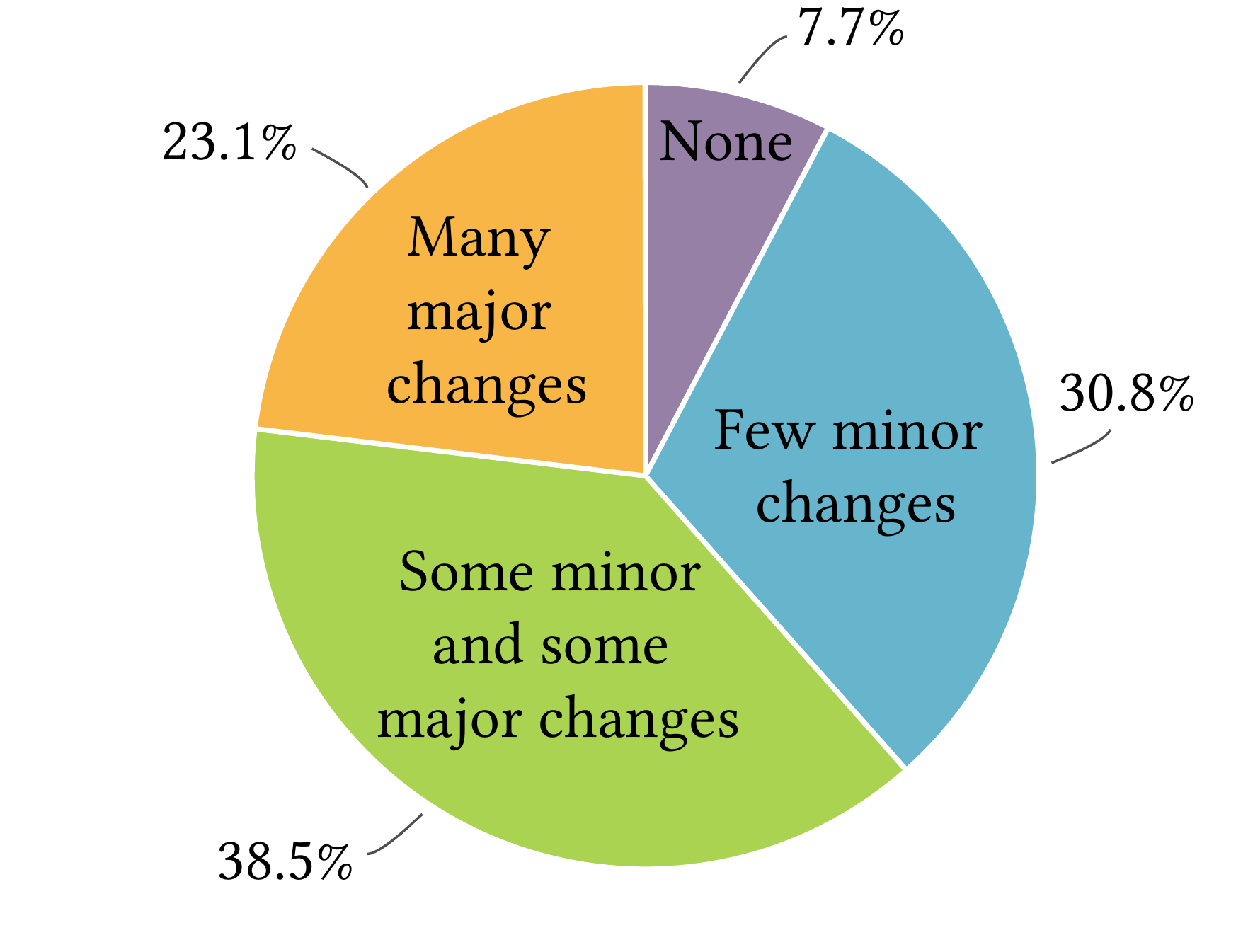}
        \label{fig:q_17}
    }%
    \subfloat[Types of changes (Q18)]{
        \includegraphics[width=0.22\linewidth]{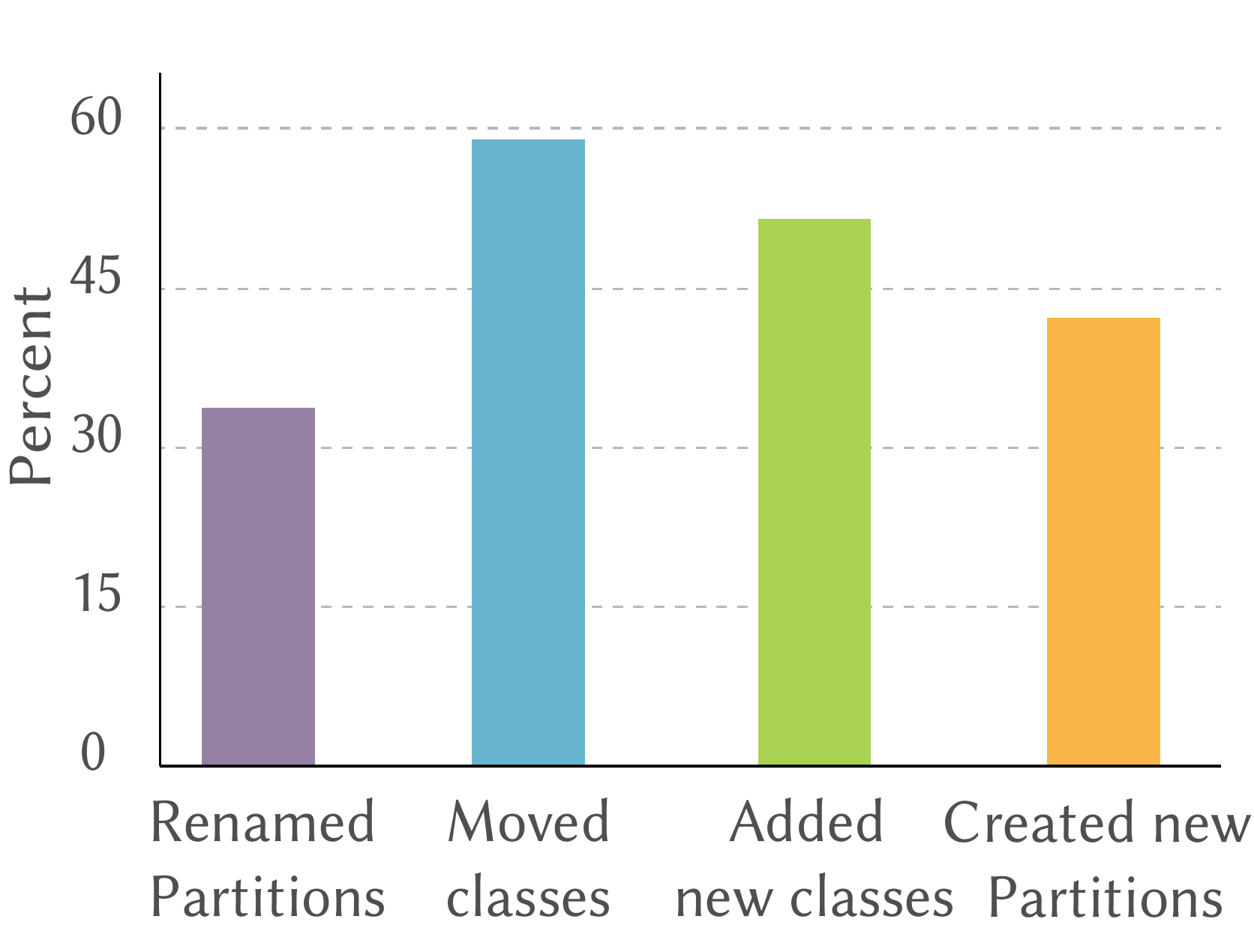}
        \label{fig:q_18}
    }%
    \subfloat[Number of partitions (Q19)]{
        \includegraphics[width=0.22\linewidth]{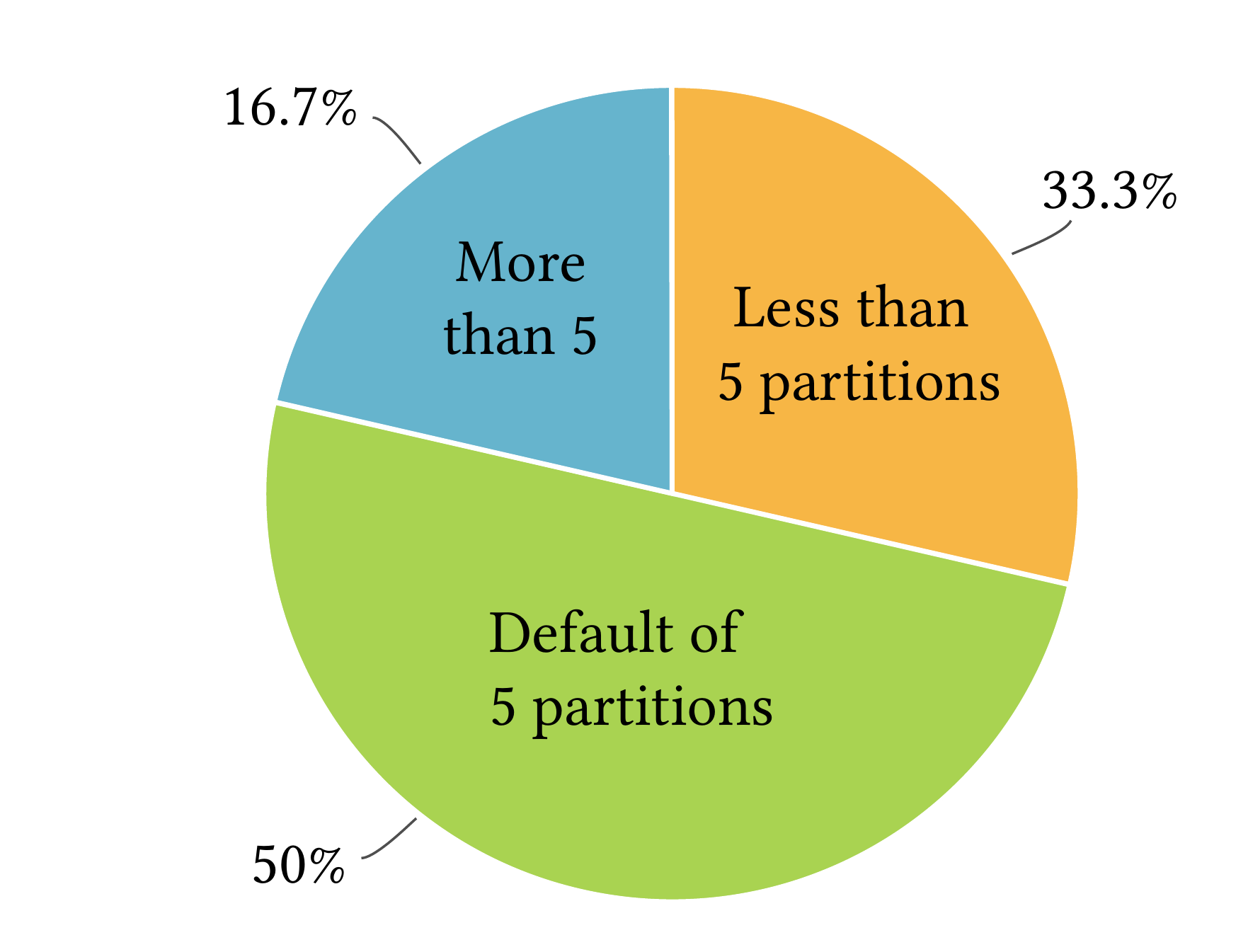}
        \label{fig:q_19}
    }%
    \subfloat[\tool does not slow down workflow (Q20)]{
        \includegraphics[width=0.22\linewidth]{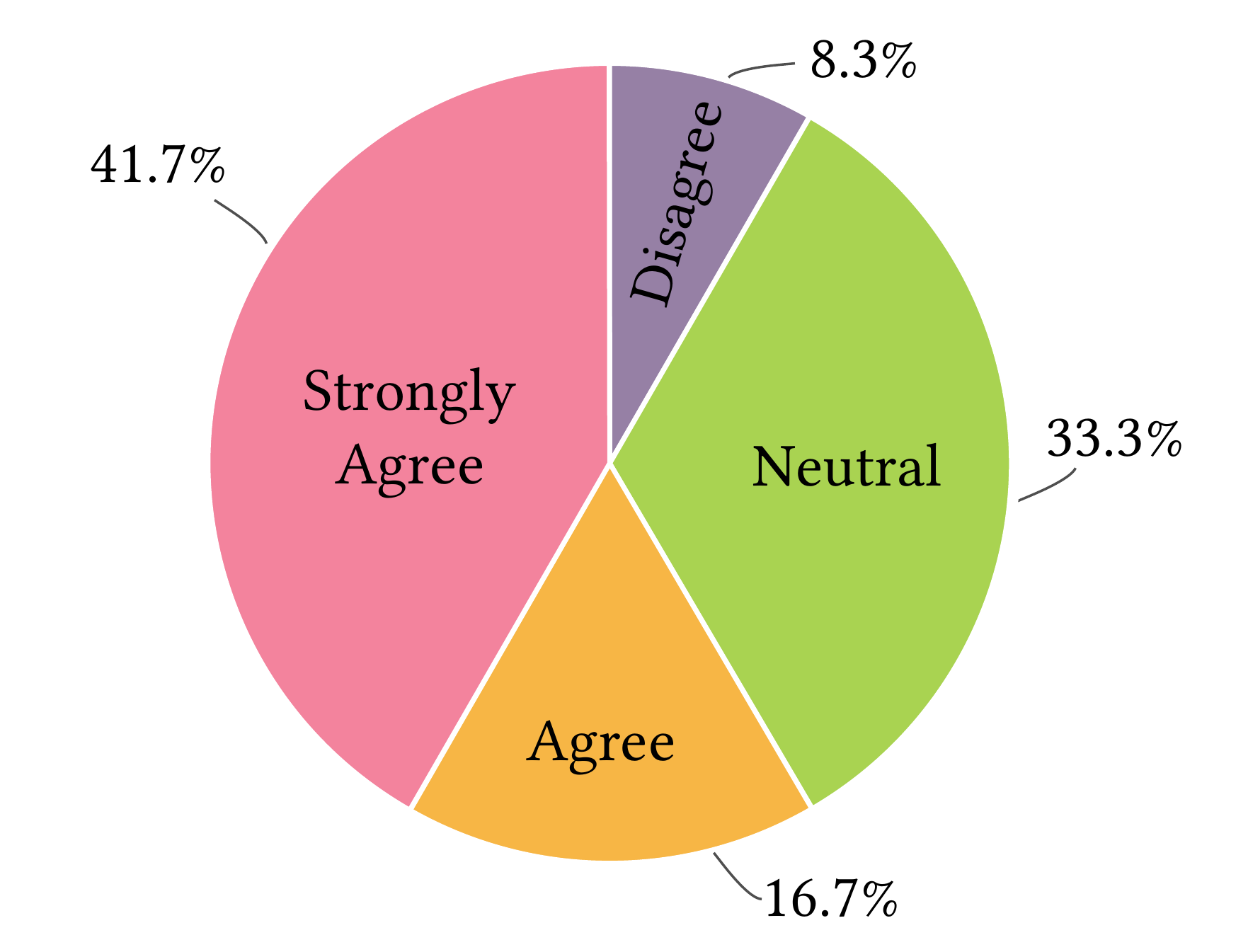}
        \label{fig:q_20}
    }%

    \caption{Survey responses from participants for Q1 to Q21 given in Table~\ref{tab1-survey-qs}.}
    \label{fig:survey_charts}
\end{figure*}

\subsubsection{Architectural Decisions}

In terms of architectural decisions, we observed the following responses. Based on Q6, most participants (57.1\%) agreed that \tool helps to implement the Strangler pattern. The response is expected considering \tool allows users to refactor their applications incrementally. The response also corroborates Fritzsch {\etal}'s \cite{Fritzsch+ICSME+2019} study where 7 of 9 cases they analyzed were re-written applying the Strangler pattern. Based on Q7, we find that most participants (41.7\%) are undecided if \tool helps them to implement DDD patterns. This partially correlates with Fritzsch {\etal} \cite{Fritzsch+ICSME+2019} findings. Though DDD has been cited frequently in the literature, only 3 of the 16 participants in Fritzsch {\etal}'s \cite{Fritzsch+ICSME+2019} study reported using it. Based on Q8, most participants (53.8\%) agree that partitions created by \tool are aligned with the business functionality of their applications.

Based on Q9, we find the following.  1) In the case of structural relations \cite{Kirby+ICPC+2021}, most participants (64.3\%) considered it extremely important. 2) In the case of semantic relations \cite{Kirby+ICPC+2021}, most participants considered it neutral (35.7\%). 3) In the case of evolutionary relations \cite{Kirby+ICPC+2021}, most participants (50.0\%) considered it extremely important.

Aside from run traces and use cases, Q10 lists other reported factors for consideration \cite{Kirby+ICPC+2021}: structural (static call graphs), semantic (class name similarity), and evolutionary relationships (change history, commit similarity, and contributor similarity). In addition, we added database interaction patterns and database transactions that are also considered important in terms of decomposing applications.  Based on Q10, we obtained the following responses from participants for each factor. 1)  For static call graphs, most participants (50\%) considered it extremely important. 2) For class name similarity, most participants (35.7\%) considered it neutral. 3) For change history, most participants (38.5\%) considered it neutral. 4) For commit similarity, most participants (42.9\%) considered it neutral. 5) For contributor similarity, most participants (42.9\%) considered it neutral. 6) For database interaction patterns, most participants (64.3\%) considered it extremely important. 7) For database transactions, most participants (64.3\%) considered it extremely important. Overall we find that participants consider database interaction patterns and database transactions as the most important factors for refactoring followed by static call graphs.

\begin{result}
\textbf{Lesson 2}: \tool is helpful in implementing the Strangler pattern \\
\textbf{Lesson 3}: Database interaction patterns and database transactions should be added to enhance \tool's partitioning strategy.
\end{result}

\subsubsection{Running \tool}

Considering running \tool, we observed the following responses. Based on Q11, most participants (50\%) agreed that \tool supports independent or mutually exclusive business functionalities. Based on Q12, we obtained the following reasons for dependencies in business functionalities: 1) strong coupling, 2) inheritance and database interactions, 3) interdependent operations, and 4) shared classes underlying technical components. Based on Q13, most participants affirmed that \tool provides a new perspective to their applications. For example, one participant responded that interactions among application classes got clearer due to the \tool's recommendations. Based on Q14, most participants (78.6\%) manually executed the use cases. The manual effort is required since several legacy monoliths may not have sufficient coverage of automated tests aligned with business functionalities. Based on Q15, we observed that most participants (54.5\%) responded that the use cases and unobserved classes align with their expectations. Most participants (54.5\%) participants responded that a gap in existing test use cases coverage was found. Most participants (63.6\%) mentioned that they found potentially dead or unreachable code using \tool.

\begin{result}
\noindent\textbf{Lesson 4}: \tool provides a new perspective to clients' applications in terms of understanding their applications business functionalities.\\
\textbf{Lesson 5}: Legacy monoliths may not have sufficient automated tests aligned with business functionalities\\
\textbf{Lesson 6}: \tool can reveal potentially dead or unreachable code
\end{result}

\subsubsection{Explainability, Configurations \& Performance}

Based on Q16, most participants responded by saying the ``explainability'' of partitions provided by \tool is valuable. Based on Q17, most participants (38.5\%) responded that they did some minor changes and some major changes. Q18 is a follow-up question based on Q17. Based on Q18, we found that most of the participants (31.8\%) suggested that they moved classes between the recommended partitions. Based on Q19, most participants (50.0\%) used the default partition value provided by \tool. Q20 is a follow-up question for Q19. Based on Q20, we found that a participant went with the default value of 5 to avoid too many microservices. Another participant responded that he/she went with a value of more than 5 since the customer was expecting more than 5 partitions. One participant responded that he/she chose a value larger than 5 since their application is relatively large with multiple domain services. The responses indicate that the partition size is dependent on the domain knowledge of applications. Based on Q21, most participants (41.7\%) agree that \tool is fast enough to generate recommendations.

\begin{result}
\textbf{Lesson 7}: Domain knowledge of an application is needed to chose the appropriate partition size.\\
\textbf{Lesson 8}: \tool's explainability of partitions in terms of use cases is valuable to users.
\end{result}

\section{Discussion}
\label{sec:summarize}

\textbf{Summary of RQ1 and RQ2}. In terms of empirical evaluation (Section~\ref{sec:evaluation}), we observed that \tool performs well across most of the metrics and applications. The BCP and NED \tool outperformed other baselines, whereas, for ICP and IFN, the performance was competitive with a slight edge over other approaches. For SM, it lost to both Bunch and MEM; however, we also observed higher SM values lead to higher NED scores. The result needs further investigation to understand the relationship between SM and other metrics. In terms of time required \tool again lost to Bunch; however, it significantly outperformed other approaches.

\textbf{Summary of RQ3}. In terms of survey (Section~\ref{sec:survey}), we observed that \tool was beneficial in several cases. 1) It helps implement the Strangler pattern; the partitions generated by \tool align with the applications' business functionality. 2) It made the interaction among classes more evident. 3) It helped users to find potentially unreachable or dead code. 4) It discovered the gap between test cases coverage. 5) It produces explainable partitions. The survey also provided further scope for improvement, such as to 1) consider static call graphs in addition to runtime traces, 2) consider database interactions and transaction patterns to improve partitioning, 3) minimize the changes required post-recommendations.

\section{Threats to Validity} 
\label{sec:threats}

Although the empirical evaluation and the survey show the effectiveness of \toolnsp, there are threats to the validity of our results. The most significant of these are threats to external validity, which arise when the observed results cannot be generalized to other experimental setups. Our evaluation included seven applications with varying use cases and code coverage. Therefore, we can draw limited conclusions on how our results might generalize to other applications, use cases, and coverage. Although our subjects have considerable variations in number and granularity of use cases and coverage achieved by the use cases, the effect of application decomposition is an aspect that requires further experimentation and investigation.

Threats to internal validity may be caused by bugs in \tool, our experimental infrastructure, and data-collection scripts. We mitigated the threat by adding validation scripts and providing appropriate error messages. For the survey, we have limited the number of participants (21) with varying degrees of job roles and experiences who completed the survey. We can address the lack of participants by creating an extensive study group to find more general results. Additionally, since \tool was generally available in January 2021, many participants did not get a chance to use it for many production applications. We think the use of \tool on a large number of production applications by survey participants could have possibly shown results favoring \tool in generating partitions for production applications.

\section{Related Work}
\label{sec:related}


In this section, first, we discuss the techniques that are most related to ours. Then, we discuss selected contributions from the software decomposition and service extraction.

\paragraph{Software Remodularization}
Microservice decomposition is a newer instance of the long-standing problem of software (re)modularization and clustering, which has seen a long line of work (\eg \cite{anquetil:1999,bavota:2010,bavota:2014,caldiera:1991,doval:1999,hutchens:1995,mahdavi:2003,maqbool:2007,mitchell:2006,mkaouer:2015,Patel+ECSMRE+2009,praditwong:2011,saeidi:2015,santos:2014,schwanke:1991,wiggerts:1997, Xiao+2005+CSMR}). We discuss select techniques from this body of work, observing that our approach, unlike the existing techniques, applies clustering on execution traces generated using functional use cases. In addition, we leverage the temporal relations as indirect call relations to generate partitions.

Commonly investigated approaches in modularization build a module dependence graph (MDG) using various types of dependence relations and then apply clustering or evolutionary algorithms to compute partitions based on different similarity metrics and objective functions.  For example, \citeauthor{doval:1999}~\cite{doval:1999} and \citeauthor{mitchell:2006}~\cite{mitchell:2006} apply genetic algorithms to the MDG to optimize a metric based on cohesion and coupling. \citeauthor{mahdavi:2003}~\cite{mahdavi:2003} investigate multiple hill climbing for software clustering.  \citeauthor{Xiao+2005+CSMR}~\cite{Xiao+2005+CSMR} consider runtime calls and associate weights with edges in the MDG. \citeauthor{bavota:2010}~\cite{bavota:2010} analyze information flowing into and out of a class via parameters of method calls; they also infer semantic information from comments and identifiers. Much of this work combines multiple goals into a single objective function, but several multi-objective formulations of modularization have been presented as well (\eg \cite{abdeen:2013,mkaouer:2015,oliveira:2012,praditwong:2011}).

\paragraph{Decomposition via Dynamic Traces}
\citeauthor{Patel+ECSMRE+2009}~\cite{Patel+ECSMRE+2009} present a decomposition technique that applies hierarchical clustering over execution traces. Their approach performs clustering over a matrix in which rows represent classes, columns represent features, and each cell has a boolean value indicating whether a class occurs in a trace. \citeauthor{Jin+TSE+2019}~\cite{Jin+2018+ICWS,Jin+TSE+2019} present a technique for identifying candidate microservices that uses execution traces collected from functional test cases. Their approach first performs function atom generation, applying hierarchical clustering based on occurrences of classes in execution traces \cite{Jin+2018+ICWS} followed by the application of a genetic algorithm to merge such atoms. \citeauthor{Alwis+ICSOC+2018}~\cite{Alwis+ICSOC+2018} propose an approach that recommends microservices at the level of class methods. For recommendations, they rely on execution traces generated from use cases and database tables. For generating partitions, they use an approach that computes subgraphs from a given graph.

\paragraph{Other Decomposition Techniques}
Several other techniques have been presented on software decomposition for microservice extraction (\eg \cite{Taibi+CLOSER+2019,Escobar+CLEI+2016,Levcovitz+CoRR+2016,Fritzsch+CoRR+2-18,Ahmadvand+REW+2016,Baresi+2017+ESOCC,Mazlami+IEEE+2017,Ren+2018+MWA,Chen+APSEC+2017,carvalho:2019}). A couple of survey papers~\cite{Fritzsch+CoRR+2-18,ponce:2019} provide an overview of recent work on this topic.

Escobar {\etal} \cite{Escobar+CLEI+2016} present a rule-based approach for clustering highly-coupled classes in JEE applications; their approach considers entity beans (representing data) and their relationships to session beans in the business tier of the application.  Levcovitz {\etal} \cite{Levcovitz+CoRR+2016} propose an approach that analyzes control flow through application tiers---from the presentation tier to the database tables to generate candidate microservices. \citeauthor{Mazlami+IEEE+2017}~\cite{Mazlami+IEEE+2017} present a graph-based clustering approach for identifying microservices. They use four different extraction strategies based on change history and developer contribution to the codebase.

Other approaches for microservice decomposition match terms in OpenAPI specifications against a reference vocabulary~\cite{Baresi+2017+ESOCC}, leverage domain-driven design and entity-relationship models~\cite{Gysel+2016+ECSOCC}, use manually constructed data-flow diagrams~\cite{Chen+APSEC+2017}, and include security and scalability requirements~\cite{Ahmadvand+REW+2016}. \citeauthor{Ren+2018+MWA}~\cite{Ren+2018+MWA} apply $k$-means clustering over combined static and runtime call information to generate microservices. \citeauthor{Taibi+CLOSER+2019}~\cite{Taibi+CLOSER+2019} apply process mining on runtime logs files to construct call graphs for partitioning.

\section{Potential Ethical Impact}
We consider the current contribution does not pose any societal or ethical impact.

\section{Datasets}
\label{sec:datasets}

We have released the datasets\footnote{\textbf{https://github.com/kaliaanup/mono2micro-fse-industry-track-2021}} for \tool and baselines. Additionally, we provide the Python-based data converters to convert \tool's dataset to the formats required by other baselines.


\section{Conclusion}
\label{sec:conclusion}

The paper provided an approach that recommends microservices from legacy applications. The approach captures and preserves the temporal relationships; it uses the relationships to group classes into disjoint partitions. Our experimental studies show the efficacy of our approach when compared with the baselines.

In the future, we plan to continue our investigation, expand the quality metrics and provide further guidance to create efficient use cases for the practitioners. We are conducting extensive verification and validation of our approach by trying it against large enterprise real-life applications in production for several years in various industry sectors. Based on the lesson learned from the survey, we plan to take the following directions: 1) Add database interaction and transaction patterns to refine \tool's recommendation. 2) Automate test case generation for legacy monoliths to generate runtime traces. 3) Automate the generation of a partition size for a legacy application. 4) Redefine NED constraints for larger applications. 5) Finally, how we can improve the explainability of partitions further.

\begin{acks}
\label{sec:ack}

We are grateful to Troy Bjerke for helping us with the survey. We would like to thank our partners at IBM Hybrid Cloud (Melissa Modjeski, Laura Scott, Dana Price, Erin Heximer) and the entire \tool development team. We are grateful to our colleagues at IBM Research Chen Lin, John Rofrano, Shivali Agarwal, Amith Singhee, Srikanth Tamilselvan, Yasuharu Katsuno, Fumiko Satoh, Nicholas Fuller, and Ruchir Puri for their valuable suggestions and feedback. Finally, we thank Julia Rubin, Evelien Boerstra, and Lisa Kirby at the University of British Columbia and Tim Menzies, Rahul Yedida, Munindar P. Singh, and Arvind Kumar at North Carolina State University for their valuable feedback on this work.
\end{acks}

\bibliographystyle{ACM-Reference-Format}
\bibliography{reference}

\end{document}